\documentclass[a4paper,10pt]{article}
\pdfoutput=1
\usepackage[table]{xcolor}
\usepackage{jcappub}
\usepackage{bbold}
\usepackage{aas_macros}
\usepackage[english]{babel}
\usepackage[utf8]{inputenc}
\usepackage{color}
\usepackage{amsfonts}
\usepackage{graphicx}
\usepackage{amssymb} 
\usepackage{etoolbox}
\usepackage{amsmath}
\usepackage{empheq}
\usepackage{cancel}
\usepackage{subfig}
\usepackage[most]{tcolorbox}
\usepackage{float}              
\usepackage{listings}
\usepackage{multirow}

\newtcbox{\mymath}[1][]{%
    nobeforeafter, math upper, tcbox raise base,
    enhanced, colframe=yellow!30!black,
    colback=yellow!30, boxrule=1pt,
    #1}

\def\be{\begin{equation}}
\def\ee{\end{equation}}
\def\bea{\begin{eqnarray}}
\def\eea{\end{eqnarray}}
\def\bean{\begin{eqnarray*}}
\def\eean{\end{eqnarray*}}

\newcommand{\code}[1]{\texttt{#1}}

\newcommand{\HH}{\mathcal H}

\newcommand{\lcdm}{\ensuremath{\Lambda}CDM}
\newcommand{\kess}{\ensuremath{k}-essence }

\definecolor{bittersweet}{rgb}{1.0, 0.44, 0.37}

\newcommand {\redt}{{ \cellcolor{red!25} }}
\newcommand {\bluet}{{ \cellcolor{blue!25}  }}

\newcommand {\gev}{{\itshape{gevolution} }}
\newcommand {\DE}{{\rm DE }}
\newcommand {\kev}{{$k$-evolution }}
\newcommand{\class}{\texttt{CLASS }}

\definecolor{somered}{rgb}{0.7, 0.06, 0.1}

\newcommand{\hMpc}[1]{\ensuremath{#1\,h^{-1}\mathrm{Mpc}}}
\newcommand{\Mpch}[1]{\ensuremath{#1\,h\mathrm{Mpc}^{-1}}}
\newcommand{\cssq}{\ensuremath{c_s^2}}

\usepackage{scalerel,tikz}
\usetikzlibrary{svg.path}
\definecolor{orcidlogocol}{HTML}{A6CE39}
\tikzset{orcidlogo/.pic={
 \fill[orcidlogocol] svg{M256,128c0,70.7-57.3,128-128,128C57.3,256,0,198.7,0,128C0,57.3,57.3,0,128,0C198.7,0,256,57.3,256,128z};
 \fill[white] svg{M86.3,186.2H70.9V79.1h15.4v48.4V186.2z}
 svg{M108.9,79.1h41.6c39.6,0,57,28.3,57,53.6c0,27.5-21.5,53.6-56.8,53.6h-41.8V79.1z M124.3,172.4h24.5c34.9,0,42.9-26.5,42.9-39.7c0-21.5-13.7-39.7-43.7-39.7h-23.7V172.4z}
 svg{M88.7,56.8c0,5.5-4.5,10.1-10.1,10.1c-5.6,0-10.1-4.6-10.1-10.1c0-5.6,4.5-10.1,10.1-10.1C84.2,46.7,88.7,51.3,88.7,56.8z};
}}
\newcommand\orcidicon[1]{\href{https://orcid.org/#1}{\mbox{\scalerel*{
\begin{tikzpicture}[yscale=-1,transform shape]
\pic{orcidlogo};
\end{tikzpicture}
}{|}}}}

\title{Parametrising non-linear dark energy perturbations}
\author[a]{Farbod Hassani~\orcidicon{0000-0003-2640-4460},}
\author[b,c]{Benjamin L'Huillier~\orcidicon{0000-0003-2934-6243},}
\author[b,d]{Arman Shafieloo~\orcidicon{0000-0001-6815-0337},}
\author[a]{Martin Kunz~\orcidicon{0000-0002-3052-7394}}
\author[e]{and Julian Adamek~\orcidicon{0000-0002-0723-6740}}

\affiliation[a]{
Universit\'e de Gen\`eve, D\'epartement de Physique Th\'eorique and CAP,
24 quai Ernest-Ansermet, CH-1211 Gen\`eve 4, Switzerland
}
\affiliation[b]{
Korea Astronomy \& Space Science Institute, Yuseong-gu, Daedeok daero 776, 34055 Daejeon, Korea}
\affiliation[c]{
Department of Astronomy, Yonsei University, 50 Yonsei-ro, Seodaemun-gu, 03722 Seoul, Korea
}
\affiliation[d]{
University of Science and Technology, Yuseong-gu 217 Gajeong-ro, Daejeon 34113,  Korea
}
\affiliation[e]{
School of Physics and Astronomy, Queen Mary University of London, 327 Mile End Road, London E1~4NS, UK
}

\emailAdd{farbod.hassani@unige.ch} 
\emailAdd{benjamin@kasi.re.kr}
\emailAdd{shafieloo@kasi.re.kr}
\emailAdd{martin.kunz@unige.ch}
\emailAdd{julian.adamek@qmul.ac.uk}

\arxivnumber{1910.01105}

\abstract{
In this paper, we quantify the non-linear effects from $k$-essence dark energy through an effective parameter $\mu$ that encodes the additional contribution of a dark energy fluid or a modification of gravity to the Poisson equation. This is a first step toward quantifying non-linear effects of dark energy/modified gravity models in a more general approach. We compare our $N$-body simulation results from $k$-evolution with predictions from the linear Boltzmann code \texttt{CLASS}, and we show that for the $k$-essence model one can safely neglect the difference between the two potentials, $ \Phi -\Psi$, and short wave corrections appearing as higher order terms in the Poisson equation, which allows us to use single parameter $\mu$ for characterizing this model. We also show that for a large $k$-essence speed of sound the \class results are sufficiently accurate, while for a low speed of sound non-linearities in matter and in the $k$-essence field are non-negligible. We propose a $\tanh$-based parameterisation for $\mu$, motivated by the results for two cases with low ($c_s^2=10^{-7}$) and high ($c_s^2=10^{-4}$) speed of sound, to include the non-linear effects based on the simulation results. This parametric form of $\mu$ can be used to improve Fisher forecasts or Newtonian $N$-body simulations for $k$-essence models.}
\begin{document}
\maketitle
\section{Introduction}
\label{sec:intro}
The accelerating expansion of the Universe is now well established based on observational results, for example from observations of the cosmic microwave background (CMB) anisotropies \cite{2016A&A...594A..13P}, type Ia supernovae \cite{2018ApJ...859..101S}, and baryon acoustic oscillations \cite{2017MNRAS.470.2617A}. Current data are in agreement with a cosmological constant as the driving force behind the acceleration. 
However, the cosmological constant suffers from severe fine-tuning issues that motivated the development of a plethora of modified gravity (MG) and dark energy (DE) models.

One of the simplest and most popular of these models is
$k$-essence\footnote{It is worth mentioning that we use the term ``$k$-essence'' to refer to a general class of models featuring either a canonical (quintessence \cite{Peebles:1998qn}) or a non-canonical kinetic term in the action.}, featuring a single scalar field minimally coupled to gravity. The $k$-essence model was originally introduced to naturally explain why the universe has entered an accelerating phase without fine-tuning of the initial conditions and the parameters and also avoiding anthropic reasoning \cite{ArmendarizPicon:2000ah}.

 Understanding the mysterious nature of DE/MG has become one of the most important unsolved problems in cosmology. Upcoming surveys  \cite{2018LRR....21....2A, 2015aska.confE..19S,4MOST:2019,Aghamousa:2016zmz} are planned with the aim of understanding this component by probing the expansion history of the Universe as well as structure formation with unprecedented precision. These surveys will put tight constraints on the cosmological parameters, including those that describe the physical properties of dark energy. While there have been many studies (for a review see e.g.\ \cite{2018LRR....21....2A}) on probing the DE/MG equation of state by observing the expansion history of the Universe, a consistent study of these theories including the non-linearities of DE/MG is not generally available. But to reach the full potential of constraining these theories, a modelling of the observables up to non-linear scales will be necessary. To this end, some of us have recently developed the $N$-body code ``$k$-evolution'' in which  we consider dark energy, in this case a $k$-essence scalar field, as an independent component in the Universe. This code is described in more detail in the companion paper \cite{kevolution}, where we also study the effect of dark matter and
gravitational non-linearities on the power spectrum of dark matter, of dark energy and of the
gravitational potential, and compare $k$-evolution to Newtonian N-body simulations.
 
Studying the non-linearities of such models in a consistent way enables us to predict the effects of DE/MG perturbations on the cosmological parameters. A full study of the $k$-essence model is particularly interesting as the dark energy perturbations become important at different scales with different amplitudes, depending on the equation of state $w$ and speed of sound  $c_s$. For example, in \cite{kevolution} we show that $k$-essence structures for low speed of sound, e.g. $c_s^2 = 10^{-7}$, can become highly non-linear at small scales and that non-linearities have a large impact on quantities like the gravitational potential power spectrum.

Providing a sufficiently precise modelling of power spectra specifically for $k$-essence dark energy is the main goal of this article. We also discuss the different sources to the Hamiltonian constraint in the presence of $k$-essence dark energy to show that the impact of $k$-essence on power spectra can be captured by the modified gravity parameter $\mu$ on all scales of interest.

In Section~\ref{sec:kess} we review the theory of $k$-essence and describe the theoretical framework. In Section~\ref{sec:res} we describe the numerical results, based on the \kev code \cite{kevolution}, a relativistic $N$-body code for clustering dark energy, and in Section \ref{sec:use} we provide basic recipes for how to use our results to improve Boltzmann and Newtonian $N$-body codes.

\section{The $k$-essence model}
\label{sec:kess}
$k$-essence theories are the most general local theories for a scalar field which is minimally coupled to Einstein gravity and involves at most two time derivatives in the equations of motion \cite{ArmendarizPicon:2000dh,Gleyzes:2014rba}. These theories are a good candidate for the late-time accelerated expansion as well as for the inflationary phase.
As a viable and interesting candidate for dark energy to be probed by future cosmological surveys, these theories are implemented in several cosmological Boltzmann and $N$-body codes, including CAMB \cite{Lewis_2000}, CLASS \cite{Lesgourgues:2011re}, \textsc{concept} \cite{Dakin:2019vnj}, \textit{gevolution} \cite{2016JCAP...07..053A} and recently in $k$-evolution \cite{kevolution}. In $k$-evolution, contrary to other codes that only consider linear dark energy perturbations, dark energy can cluster like the matter components which enables us to study the effect of dark energy non-linearities in a consistent way.

In $k$-essence theories the Lagrangian is written as a general function of the kinetic term and the scalar field, $P(X,\varphi)$.
We consider the Friedman-Lema\^{i}tre-Robertson-Walker (FLRW) metric in the conformal Poisson gauge to study the perturbations around the homogeneous universe.  
\be
ds^2 = a^2(\tau) \Big[ - e^{2 \Psi} d\tau^2 -2 B_i dx^i d\tau  + \big( e^{-2 \Phi} \delta_{ij} + h_{ij}\big)  dx^i dx^j \Big]  \;,
\ee
where $\tau$ is conformal time, $x^i$ are comoving Cartesian coordinates, $\Psi$ and $\Phi$ are respectively the temporal and spatial scalar perturbations, and $B_i$ and $h_{ij}$ are the vector and tensor perturbations. Using the scalar-vector-tensor (SVT) decomposition we can recover the 4 scalar, 4 vector, and 2 tensor degrees of freedom in the metric (from which after fixing the gauge two scalar and two vector degrees of freedom are removed), which we are going to use to obtain the equations of motion for the perturbations. 
Our notation and the SVT decomposition are briefly discussed in Appendix~\ref{svt_decom}.

The full action in the presence of a $k$-essence scalar field as a dark energy candidate reads
\be
S=\frac{1}{16 \pi G_N} \int \sqrt{-g} R d^4 x + \int \sqrt{-g}  \mathcal{L}_{\DE} d^4 x + \int \sqrt{-g}  \mathcal{L}_m d^4 x ~, \label{eq:action}
\ee
where $G_N$ is Newton's gravitational constant, $g$ is the determinant of the metric, $R$ is the Ricci scalar, $\mathcal{L}_{\rm DE} = P (X, \varphi )$ is the general $k$-essence Lagrangian in which $\varphi$ is the scalar field perturbation, $X=-\frac{1}{2} g^{\mu \nu} \partial_{\mu} \varphi \partial_{\nu} \varphi$ is the kinetic term, and $\mathcal{L}_m$ is matter Lagrangian. 

Variation of the action with respect to scale factor $a(\tau$) results in an equation for the evolution of the scale factor (Friedmann equation),
\be
\frac{3}{2} \mathcal{H}^{2}=-4 \pi G_N a^{2} \bar{T}_0^0,
\ee
where $\mathcal{H} = a'/a$ and the prime here denotes the derivative with respect to conformal time. $\bar{T}_0^0$ is the background stress-energy tensor. The full stress-energy tensor including matter (cold dark matter, baryons and radiation) and $k$-essence is defined as 
\be 
T^{\mu \nu} \equiv \frac{2}{\sqrt{-g}} \frac{\delta \mathcal{L}_{\rm DE}}{\delta g_{\mu \nu}} +\frac{2}{\sqrt{-g}} \frac{\delta \mathcal{L}_{\rm m}}{\delta g_{\mu \nu}} ~.
\ee
We can parametrise the stress-energy tensor of a fluid with three parameters, namely the equation of state $w=\bar{p}/\bar{\rho}$, the squared speed of sound $c_s^2$, given in the fluid rest-frame through $\delta p = c_s^2 \delta \rho$, and the anisotropic stress $\sigma$. For \kess  both $w$ and $c_s^2$ can vary as a function of time, while $\sigma=0$ \cite{kevolution}. {However, in this work, we take $w$ and $c_s^2$ to be constant.}
On the other hand the divergence of the $k$-essence stress-energy tensor gives the equation for $k$-essence density perturbations through the continuity\footnote{In the companion paper \cite{kevolution} we show that the Euler and continuity equations are equivalent to the scalar field equation in the weak field regime.} equation,
\be
\delta'_{\rm DE} = -(1+w) \big(\partial_i v^i_{\rm DE}- 3 \Phi' \big) -3 \mathcal{H}  \bigg(\frac{\delta p _{\rm DE}} {\delta \rho_{\rm DE}} -w\bigg) \, \delta_{\rm DE} + 3  \Phi'  \bigg( 1+ \frac{\delta p_{\rm DE} } {\delta \rho_{\rm DE} }  \bigg) \, \delta_{\rm DE}+ \frac{1+w}{\rho} v^i_{\rm DE} \partial_i  \big(3\Phi - \Psi \big) \;,
\ee
where $\delta_{\rm DE}$ is the $k$-essence density contrast and $v^i_{\rm DE}$ is the velocity perturbation of $k$-essence. In this form of the continuity equation we have included short wave corrections that are discussed in more detail later in this section.

We note that in Newtonian gauge we have the relation $\delta p = c_s^2 \delta\rho + 3 \mathcal{H} (c_s^2 - c_a^2) \bar\rho (1+w) \theta/k^2$, where we have introduced the adiabatic speed of sound $c_a^2 = \bar\rho'/ \bar p'$.

The variation of the action with respect to {the lapse perturbation} $\Psi$, in the weak field approximation, results in the {Hamiltonian constraint} \cite{Adamek:2017uiq},
\be 
\nabla^2 \Phi = 3 \mathcal{H} \Phi^{\prime} + 3 \mathcal{H}^{2}\Psi + \frac{1}{2} \delta^{i j} \Phi_{, i} \Phi_{, j} + 4 \pi G_N a^{2} (1- 2 \Phi) \sum _X \bar{\rho}_X \delta_X  ~, \label{eq:Poisson}
\ee
where {$\delta_X=\delta\rho_X/\bar{\rho}_X$} is the Poisson-gauge density contrast for each species. We usually split the total density perturbation $\bar\rho\delta$ into the contribution from the different species that cluster, in our case cold dark matter, baryon, radiation and the $k$-essence scalar field:
\be
\bar{\rho}\delta = \rho_{\rm cdm} \delta_{\rm cdm} + \rho_{\rm DE} \delta_{\rm DE} + \rho_b \delta_b + \rho_r \delta_r \, ,
\ee
where cdm, $b$, and DE respectively stand for cold dark matter (CDM), baryons, and $k$-essence. The last contribution is due to relativistic species (radiation and neutrinos) that we will neglect from now on as we are interested in late times. This does however have to be taken into account when going to high redshift, e.g.\ when considering the CMB.
Moreover we define the short-wave corrections $\rm S$ and relativistic terms $\rm R$ in the {Hamiltonian constraint} equation as follows,
\begin{eqnarray}
\rm{R}(\vec x,\tau) &\equiv& 3 \mathcal{H} \Phi^{\prime} + 3 \mathcal{H}^{2}\Psi ~,  \label{eq:sw}\\
\rm{S} (\vec x,\tau) &\equiv&  \frac{1}{2} \delta^{i j} \Phi_{, i} \Phi_{, j} -8 \pi G_N a^{2}  \Phi \sum _X \bar{\rho}_X \delta_X  ~.  \label{eq:re}
\end{eqnarray}
The relativistic terms {\textbf{$\rm {R}(\vec{x},\tau)$}} become important on large scales where $k \sim \mathcal{H}$. The short wave corrections $\rm {S}(\vec{x},\tau)$, on the other hand, are due to the weak field scheme where we allow matter and $k$-essence densities to become fully non-linear i.e.
\be 
\delta_m \sim \delta_{\rm DE}\sim \mathcal{O}(1) ~,
\ee
while the metric perturbations remain small. As a result in this scheme we can have highly dense $k$-essence and matter structures while the metric is still FLRW with small perturbations. More detailed discussions on the weak field approximation are found in \cite{kevolution,2016JCAP...07..053A}

\section{Numerical results}
\label{sec:res}

\subsection{The \kev code}
\label{sec:kevcode}

\kev \cite{kevolution} is a relativistic $N$-body code based on \gev \cite{2016JCAP...07..053A,Adamek:2015eda}.
The full sets of non-linear relativistic equations, six Einstein's equations $G_{\mu \nu} = 8 \pi G T_{\mu \nu}$ as well as the scalar field equation (linearised in the $k$-essence field variables) are solved on a Cartesian grid with fixed resolution \cite{Adamek:2017uiq, kevolution} to update the particle positions and velocities. 
While in $k$-evolution the dark energy component is considered as an independent element whose equation of motion is fully coupled to the non-linear matter dynamics solved in the code, in \gev it is not treated as an independent component and only the respective linear solution from the Boltzmann code \class is used to model dark energy perturbations. 
The effects of non-linear clustering of the $k$-essence scalar field on matter and gravitational potential power spectra are studied in \cite{kevolution}. Also the effect of $k$-essence on the turn-around radius is studied using \kev in \cite{Hansen:2019juz}. 

In order to probe the non-linearities of both matter and $k$-essence scalar field, we combined the data from two simulations with $N_\text{grid} = 3840^3$ with two different resolutions: one with $L = \hMpc{9000}$ and one with a physically smaller box with $L=\hMpc{1280}$, corresponding to respectively \hMpc{2.3} and \hMpc{0.33} length resolution. Furthermore, to study the relativistic terms which become important at large scales we also use a much lower spatial resolution simulation with $N_\text{grid} = 3840^3$, $L=\hMpc{90000}$ corresponding to \hMpc{23.43} length resolution. In all of the figures, we remove the data with wavenumbers larger than 1/7 of the Nyquist frequency of the simulation to minimize any finite resolution effect. Moreover, by having different simulations with overlapping windows of wavenumbers we are able to test the convergence of the simulations and thus have control over the errors coming from finite resolution and finite box size (cosmic variance).

In our studies we consider $w= -0.9$ which gives accelerated expansion and is compatible with the current observational data \cite{2018LRR....21....2A} and two cases for the speed of sound, namely $c_s^2=10^{-4}$ and $c_s^2=10^{-7}$. These two cases are interesting as their respective sound horizons correspond to linear and non-linear scales. In fact, for the case with $c_s^2=10^{-4}$ the dark energy perturbations decay significantly at the quasi-linear and non-linear scales as the sound horizon wavenumber is about $k_{s} \approx \Mpch{0.03}$ and we do not expect to see large difference between the $k$-evolution scheme and when dark energy is treated linearly.  In the case with $c_s^2=10^{-7}$ the sound horizon for dark energy is at a much higher wavenumber, $k_{s} \approx \Mpch{1}$, and we are able to see the dark energy perturbations impact on the other quantities. More details about the results of the two speeds of sound can be found in \cite{kevolution}.

\subsection{Sources to the Hamiltonian constraint }
In this subsection we discuss the different sources to the Hamiltonian constraint and according to the numerical results we argue that in the $k$-essence theories one can safely neglect short-wave corrections and also that the gravitational potential difference $\Phi- \Psi$ is negligible.

\begin{figure}%
 \includegraphics[scale=0.5]{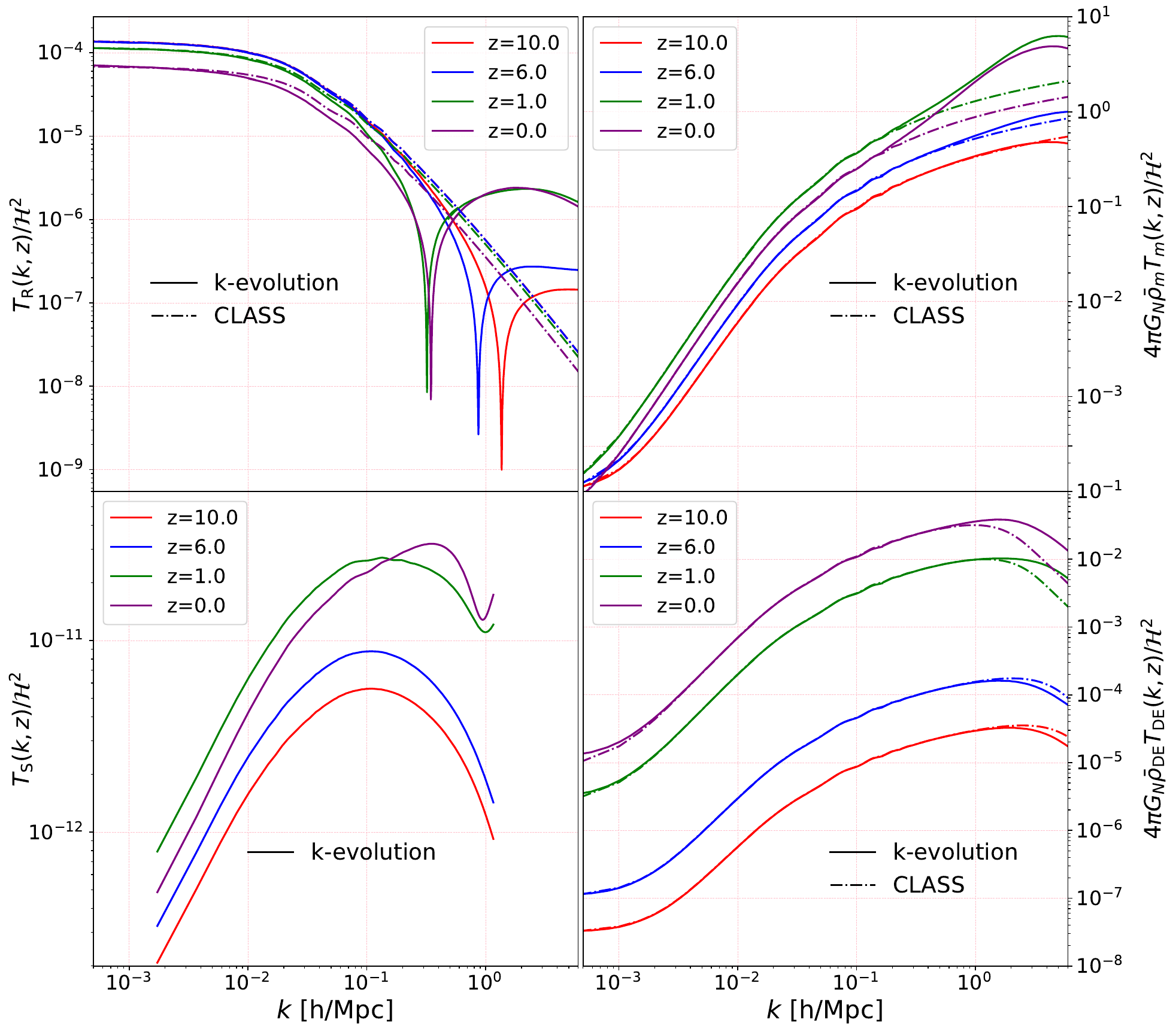} 
    \caption{Sources to the Hamiltonian constraint, normalized to $\HH^2$, in terms of wavenumber at different redshifts, from \kev in solid lines and from \class in dashed lines, for a $k$-essence speed of sound of $c_s^2=10^{-7}$. For the figures on the right, the y-axis is shown on the right. The x-axis is common between each column of figures. The figure on the left bottom, showing the short-wave corrections, is obtained using a simulation with $N_\text{grid} = 3840^3$ and box size $L = \hMpc{9000}$, while in the other three figures the results are obtained combining three simulations with $N_\text{grid} = 3840^3$ and box sizes $L = \hMpc{90000}$, $L = \hMpc{9000}$ and $L = \hMpc{1280}$.  }%
    \label{fig_powers_RE}%
\end{figure}

In Fig.~\ref{fig_powers_RE} we compare all the terms to the Hamiltonian constraint \eqref{eq:Poisson}; $T_X$ in the figures refers to $\sqrt{\langle X X^* \rangle}$ in Fourier space. We also divide these terms by $\mathcal {H}^2$ to make them dimensionless. In the top-left plot the contribution from relativistic terms, from the \kev code which is introduced in Section \ref{sec:kevcode} is shown in solid lines and from the linear Boltzmann code \class \cite{2011JCAP...07..034B} in dash-dotted lines. These terms are the main contributions to the Hamiltonian constraint at large scales and high redshifts,  as the other three terms decay at large scales. We note that the \kev and \class predictions for the relativistic terms start to differ already at relatively large scales, which comes from the difference between $\Phi'$ power spectra in these codes which is discussed in details in the companion paper \cite{kevolution}.

In the top-right figure the contribution from matter (baryons and cold dark matter) is shown. This term is the main contribution at small scales compared to the other terms, as matter perturbations dominate at those scales. The bottom-left figure  shows the contribution from the short wave corrections to the Hamiltonian constraint, which is  negligible. 
In the bottom-right figure the contribution from $k$-essence for $c_s^2=10^{-7}$ is shown. The contribution from this term peaks around the sound-horizon scale $k_{s} \approx 1 h/$Mpc. These plots allow to compare the relative contribution of each term in the Hamiltonian constraint as a function of scale and redshift; we should however not forget that the power spectra also contain contributions from cross terms that we are not going to discuss in this paper.
 
Variation of the action with respect to $\Phi_h$ (the scalar part of {$\delta g_{ij}$}) defined in Eq.~(\ref{eq:beta}) leads to a constraint equation for $\Phi -\Psi$,
\be 
\nabla^{4} (\Phi - \Psi)-\left(3 \delta^{i k} \delta^{j l} \frac{\partial^{2}}{\partial x^{k} \partial x^{l}}-\delta^{i j} \nabla^2 \right) \Phi_{, i} \Phi_{, j}=4 \pi G_N a^{2}\left(3 \delta^{i k} \frac{\partial^{2}}{\partial x^{j} \partial x^{k}}-\delta_{j}^{i} \nabla^2\right) T_{i}^{j} \, ,
\ee
where $\nabla^4 \doteq \delta^{ij}  \delta^{lm} \partial_i \partial_j  \partial_l \partial_m $. In this expression $\Phi -\Psi$ is sourced by the anisotropic part of the stress-energy tensor and a short-wave correction term. In first order perturbation theory, and neglecting radiation perturbations, we have $\Phi=\Psi$. Short-wave corrections and also anisotropic pressure generation in dark matter \cite{Ballesteros:2011cm} and $k$-essence \cite{kevolution} lead to a non-zero $\Phi- \Psi$. Contrary to the case of the Hamiltonian constraint, the contribution of short-wave corrections is of relative importance here, in particular at large scales. {In absolute terms these higher-order effects are however expected to be small. To quantify the difference between the two potentials we measure $\sqrt{\mathcal{P}_{\Phi - \Psi}}/\sqrt{\mathcal{P}_{\Phi }}$ from our simulations (where $\mathcal{P}_{\Phi - \Psi} $ is the dimensionless power spectrum of $\Phi - \Psi$), shown as solid lines in Fig.\ \ref{fig:anisotropies}. For comparison, the dashed lines show the same quantity generated from \texttt{CLASS} {where it is solely due to radiation perturbations}. On super-horizon scales, the contribution from radiation perturbations is larger than contribution from non-linearities, while in the quasi-linear regime the dominant contribution comes from the non-linearities. Both are however indeed very small and can be safely neglected at intermediate and small scales.

\begin{figure}%
     \centering
    \includegraphics[scale=0.5]{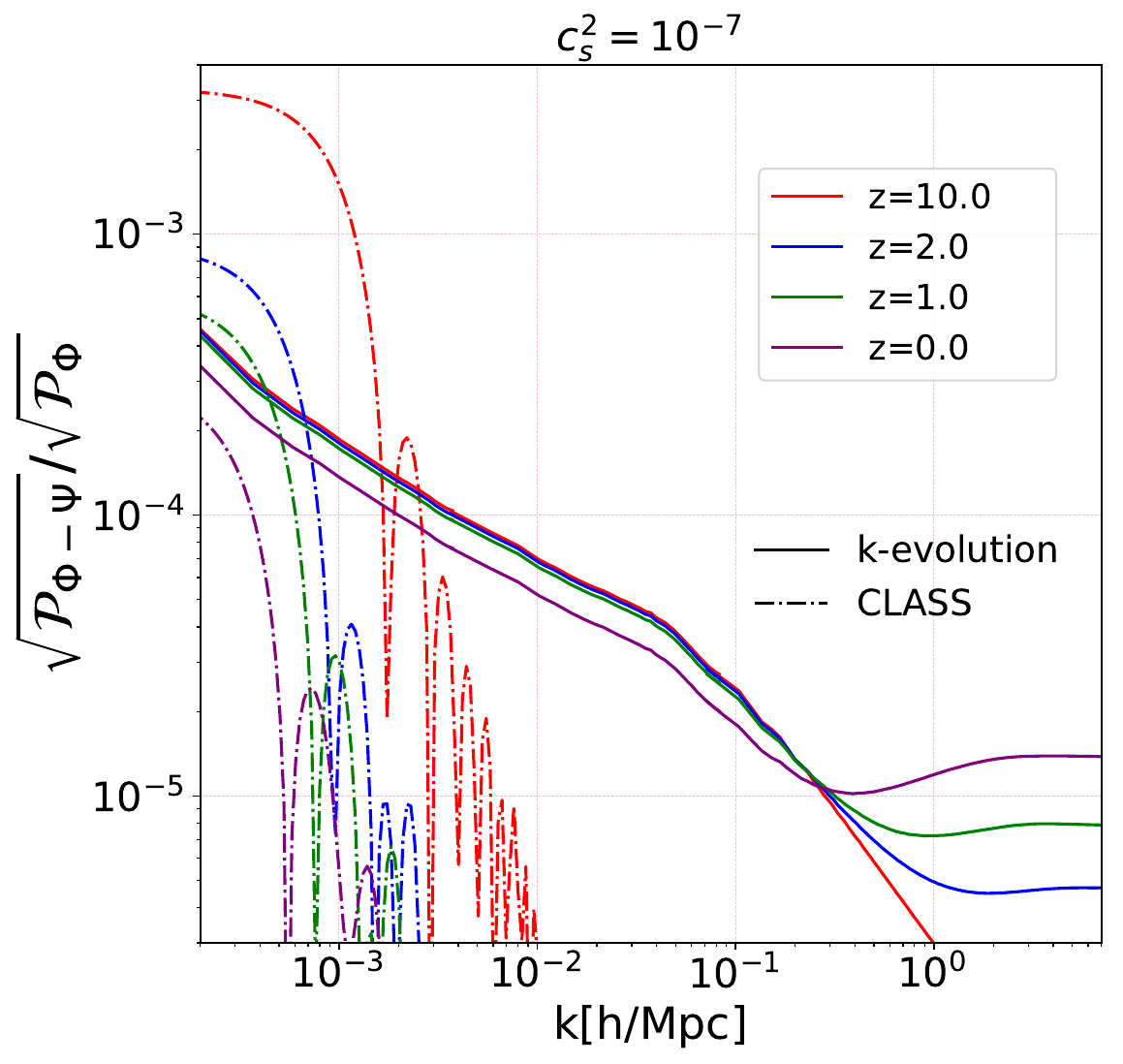}
        \caption{The quantity $\sqrt{\frac{\mathcal{P}_{\Phi - \Psi}}{\mathcal{P}_{\Phi }}}$ from \kev (solid lines) and from \class (dashed lines) at different redshifts as a function of wavenumber $k$. In \kev this quantity is non-zero due to the non-linearities in matter, $k$-essence and short-wave corrections, while in \class it is generated due to the radiation perturbations which oscillate and decay in k. The results are obtained combining three simulations with $N_\text{grid} = 3840^3$ and box sizes $L = \hMpc{90000}$, $L = \hMpc{9000}$ and $L = \hMpc{1280}$.} 
     \label{fig:anisotropies}%
 \end{figure}
Variation of the action with respect to the {shift perturbation} results in {the momentum constraint},
\be 
-\frac{1}{4} \nabla^2 B_{i}-\Phi_{, i}^{\prime}-\mathcal{H} \Psi_{,i}=4 \pi G_N a^{2} T_{i}^{0} \label{eq:vector}
\ee
where in Poisson gauge (\ref{eq:gauge}), $B_i$ is divergence-less or transverse i.e. $\delta^{ij}  {B}_{i,j}=0$. So the divergence of  Eq.~(\ref{eq:vector}) reads,
\be 
-\nabla^2 \big(\Phi^{\prime}+\mathcal{H} \Psi \big)=4 \pi G_N a^{2} \partial^i T_{i}^{0} \,. \label{eq:divergence}
\ee
{If the stress-energy can be split into contributions from independent constituents, we can define the velocity divergence $\theta_X$ for each constituent such that}
 \be 
\partial^i T_{i}^{0}  = \sum_X \bar{\rho}_X \big(1+w_X \big) \theta_X  {= \theta_\mathrm{tot} \sum_X \bar{\rho}_X \big(1+w_X \big)}\,.
\ee
{This definition of $\theta_X$ coincides with the linear velocity divergence in the case where the constituent can be described by a fluid, but it generalises to situations where this is no longer the case. We can now see that} the relativistic terms in the {Hamiltonian constraint} Eq.~\eqref{eq:Poisson} {can be related to a different choice of density perturbation},
\be 
\nabla^2 \Phi =  \frac{1}{2} \delta^{i j} \Phi_{, i} \Phi_{, j} +4 \pi G_N a^{2} (1- 2 \Phi) \sum _X \bar{\rho}_X \Delta_X  ~, \label{eq:Poisson2}
\ee
where {$\Delta_X = \delta_X - 3 \mathcal{H} (1+w_X) \nabla^{-2}\theta_\mathrm{tot}$  is the comoving density contrast.} 
{Eq.~(\ref{eq:Poisson2}) is the Poisson equation dressed with short-wave corrections, which are the leading higher-order weak-field terms. Neglecting the (small) short-wave terms,  the equation is linear even if the perturbations in the matter fields are large\footnote{{This also shows that in General Relativity the Poisson equation holds effectively on {\it all} scales, right down to milli-parsec scales where we may start to encounter effects from the strong-field regime of supermassive black holes.}}, and one can pass to Fourier space. Modifications of gravity where the gravitational coupling depends on time and scale can then be parametrised by introducing a function $\mu(k,z)$ such that
\be
-k^2 \Phi = 4 \pi G_N a^2 \mu(k,z) \sum _X \bar{\rho}_X \Delta_X\,,
\ee
where the choice $\mu(k,z)=1$ restores standard gravity. Furthermore, if one chooses to interpret the dark energy perturbations as a modification of gravity, the sum on the right-hand side would exclude $X = \mathrm{\rm DE}$. Such an interpretation makes sense if the dark energy field is not coupled directly to other matter, so that it cannot be distinguished observationally from a modification of gravity (e.g.\ \cite{Kunz:2006ca}).

Adopting this interpretation we can define an effective modification $\mu(k,z)$ as
\be
\mu(k,z)^2 = \frac{k^4 \left\langle\Phi \Phi^\ast\right\rangle}{\left(4 \pi G_N a^2 \bar{\rho}_m\right)^2 \left\langle \Delta_m \Delta_m^\ast\right\rangle}\,, \label{eq:mu_1}
\ee
where $\bar{\rho}_m = \bar{\rho}_\mathrm{cdm} + \bar{\rho}_b$ and $\bar{\rho}_m \Delta_m = \bar{\rho}_\mathrm{cdm} \Delta_\mathrm{cdm} + \bar{\rho}_b \Delta_b$. Since our simulations are carried out in Poisson gauge they internally use $\delta_m$ and do not compute $\Delta_m$ directly. However, the difference between the two quantities is only appreciable at very large scales where $\theta_\mathrm{tot}$ is given by its linear solution. For the purpose of computing $\mu(k,z)$ from simulations we therefore write,
\be
\Delta_m \simeq \delta_m \left(1 + \frac{3 \mathcal{H} T^\theta_\mathrm{tot}(k,z)}{k^2 T^\delta_m(k,z)}\right)\,,
\ee
where $T^\theta_\mathrm{tot}(k,z)$ and $T^\delta_m(k,z)$ are the linear transfer functions of $\theta_\mathrm{tot}$ and $\delta_m$, respectively. These can be computed with a linear Einstein-Boltzmann solver like \texttt{CLASS}.
}

\subsection{Linear versus non-linear $\mu(k,z)$}

In this subsection we show the $\mu(k,z)$ function obtained from $k$-evolution, which includes non-linearities in matter and $k$-essence as well as relativistic and short-wave corrections. We compare $\mu(k,z)$ from \kev with the results from the linear Boltzmann code \class \cite{2011JCAP...07..034B}, and with results from {\it gevolution} \cite{kevolution} and \class with Halofit \cite{Takahashi:2012em}. 

In the two top panels of Fig.~\ref{figmu}, the results from \kev and \class for two different speeds of sound are compared at different redshifts. In the two bottom panels $\mu(k,z)$ from $k$-evolution, \gev and \class (linear and with Halofit) at $z=0$ are shown.

\begin{figure}[t]
    \centering
    \subfloat[]{{\includegraphics[scale=0.36]{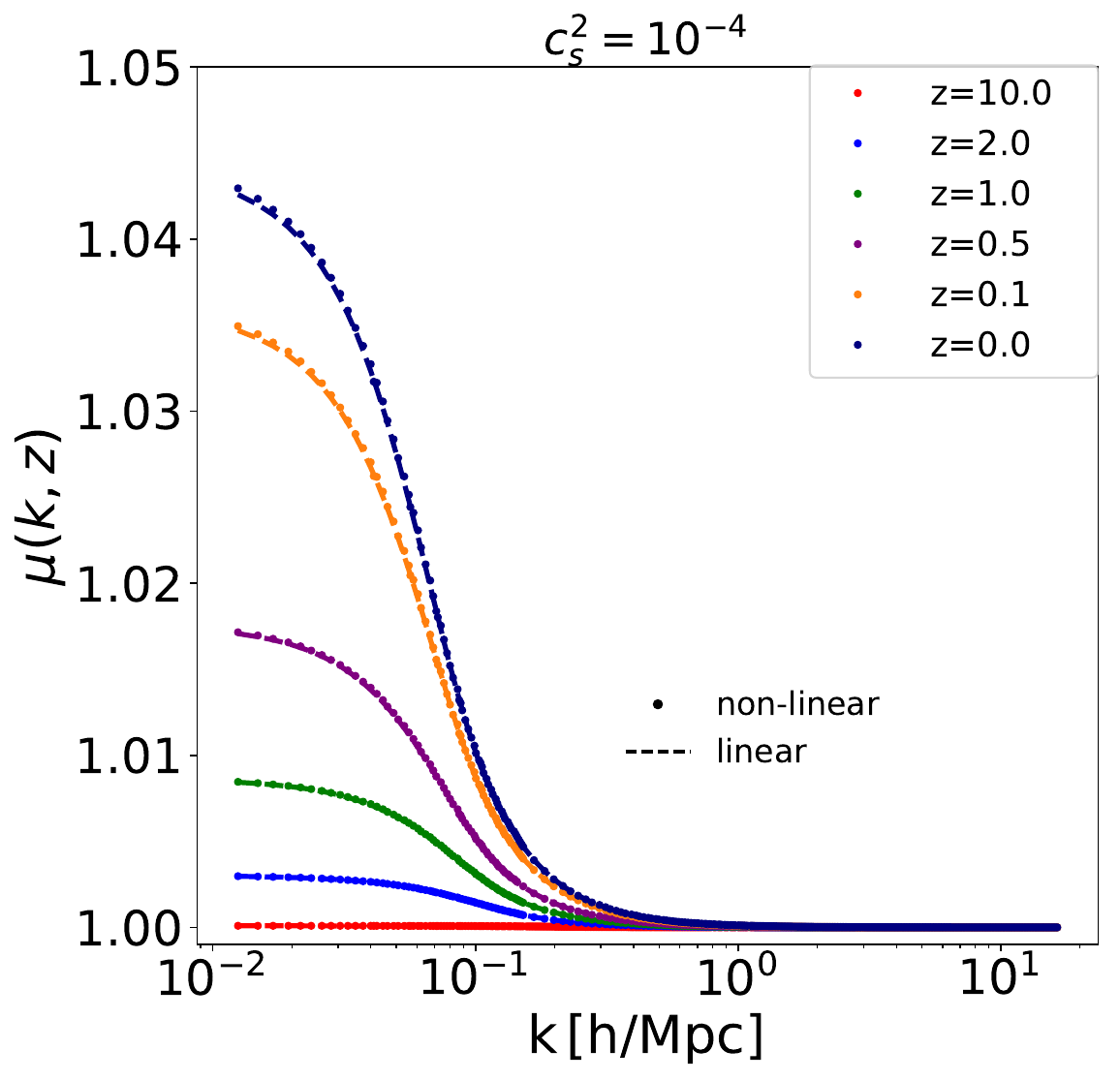} }}%
    \subfloat[]{{\includegraphics[scale=0.36]{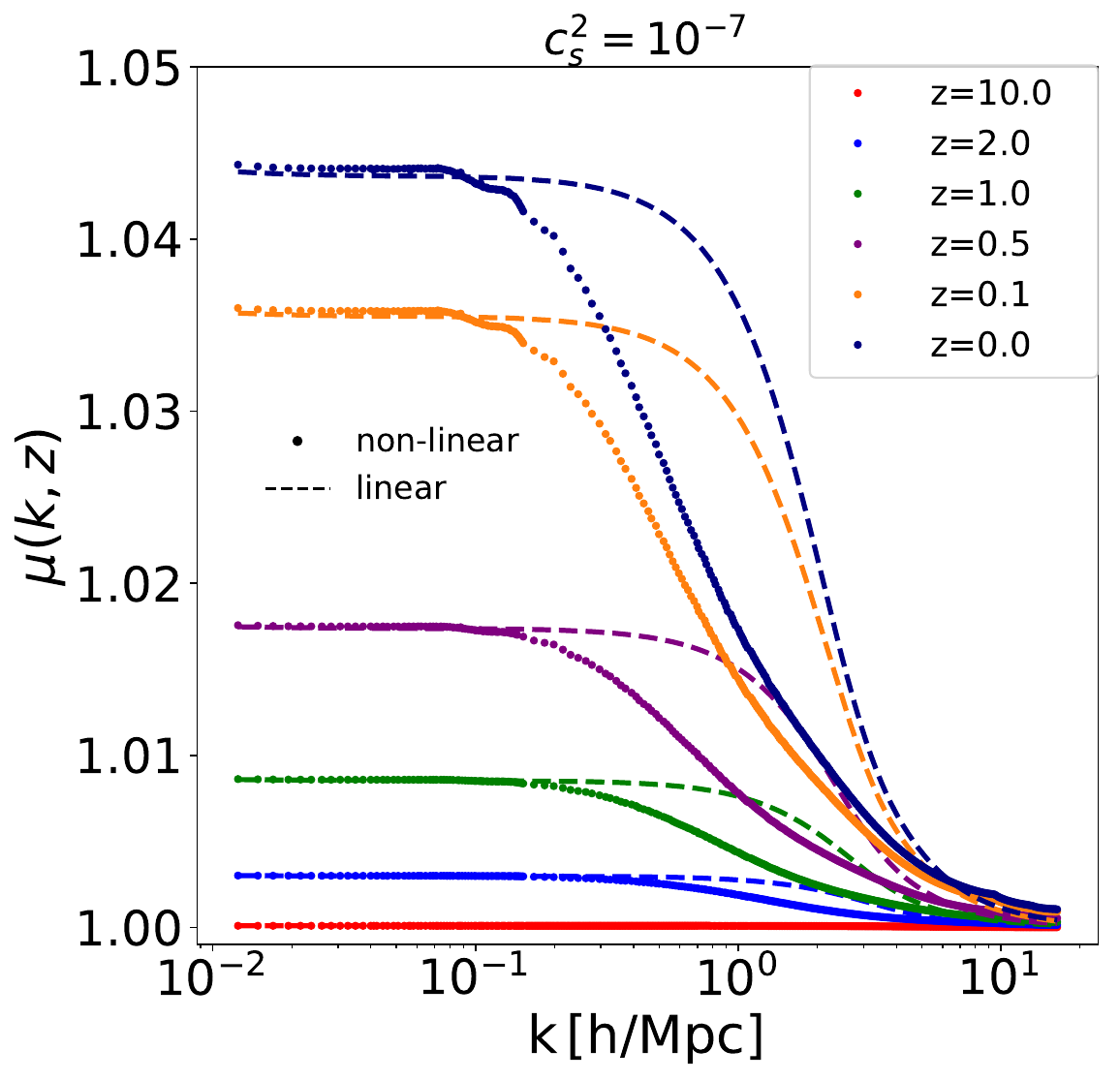} }}\\%
     \subfloat[ ]{{\includegraphics[scale=0.36]{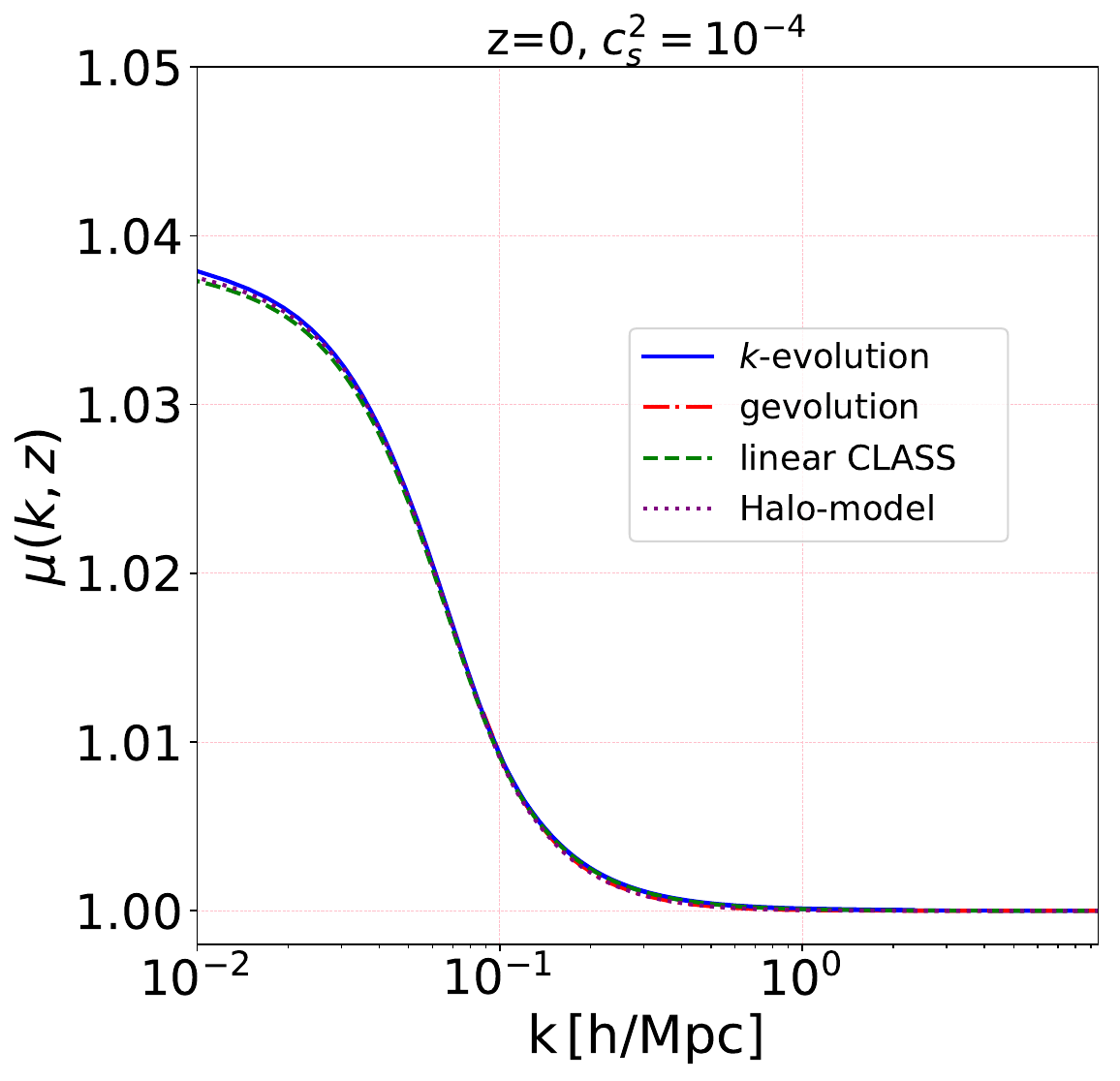}}}%
     \subfloat[ ]{{\includegraphics[scale=0.36]{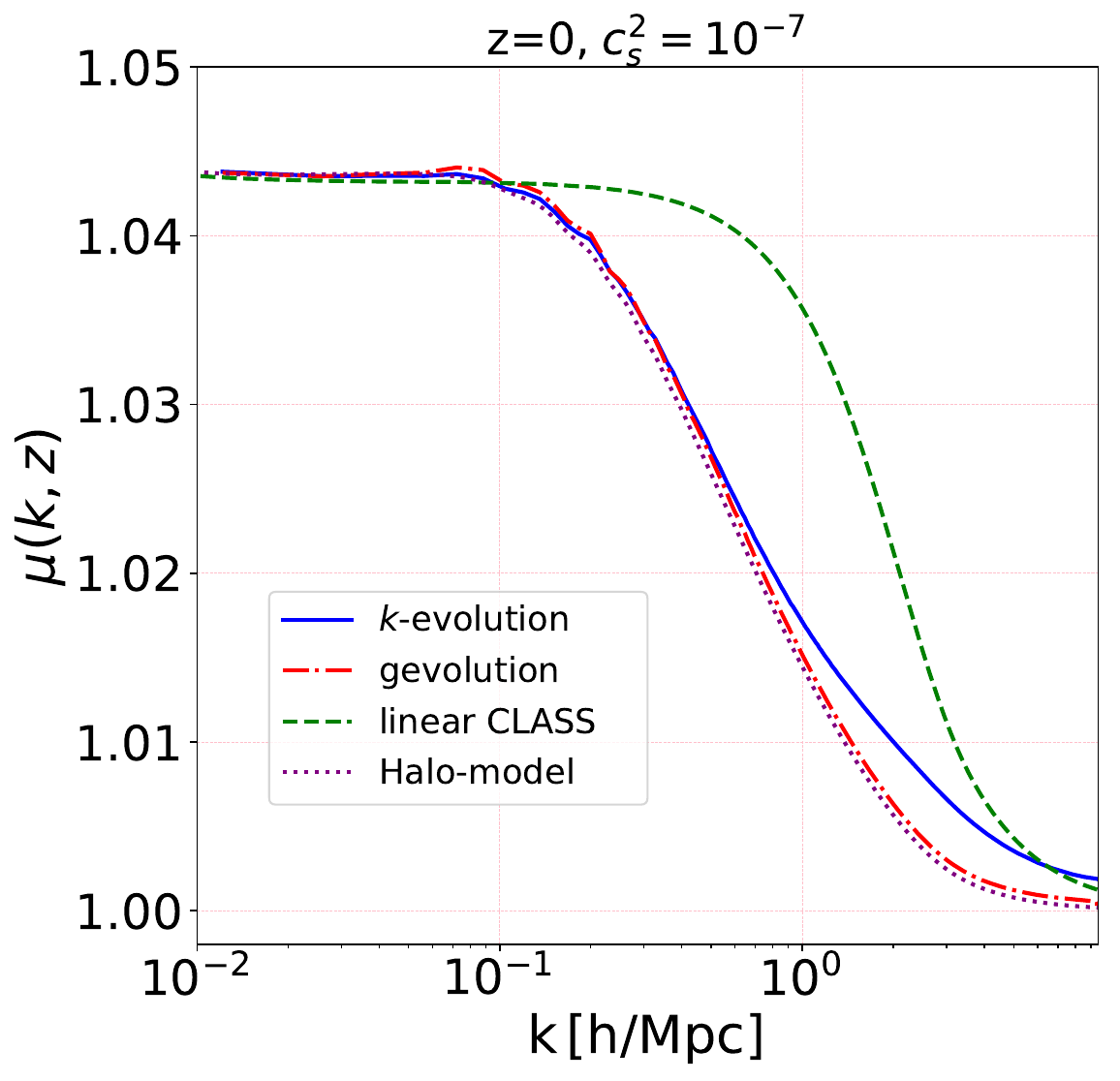} }}%
        \caption{$\mu(k,z)$ from four different simulations and for two speeds of sound are compared. The left plots depict the case $c_s^2=10^{-4}$: all the curves agree well which shows that for  high speeds of sound we can trust even linear codes. On the right we see the situation for $k$-essence with speed of sound $c_s^2=10^{-7}$, where differences are clearly visible. In the top panels \kev is compared with \class at different redshifts, while in the bottom figures we compare $\mu(k,z)$ from {\it gevolution}, \kev and \class with and without Halofit, at $z=0$. The results are obtained using two simulations with two different box sizes $L = \hMpc{9000}$ and $L = \hMpc{1280}$ and $N_\text{grid} = 3840^3$.} 
     \label{figmu}%
\end{figure}

In \lcdm, where there are no dark energy perturbations, or modifications of gravity, we would have $\mu = 1$ on all scales and at all times. In Fig.~\ref{figmu} we see that due to the $k$-essence perturbations we have $\mu > 1$ at large scales, while we recover the \lcdm\ limit on small scales or at early times. The maximum deviation from $\mu=1$ is less than 5\% for our choice of model, therefore the effect is small {but not negligible}.
The reason why we expect to recover GR at high redshifts is that the DE/MG starts to dominate at low redshifts which is well supported by observations. In our model this is included by choosing a constant $w=-0.9$ close to $-1$, in which case the ratio of dark energy to dark matter density scales like $a^{-3w}$, ensuring that the dark energy quickly becomes sub-dominant in the past.

At high wavenumbers, the $k$-essence perturbations are suppressed due to the existence of a sound-horizon, roughly at the comoving wavenumber $k = c_s \mathcal{H}$. This is highly desirable as gravity is very well tested on small scales, so that models that lead to significant changes on solar system scales are ruled out by observations. {The scaling of the sound horizon with $\cssq$ also explains why the transition from $\mu>1$ to $\mu=1$ occurs on smaller scales for lower speeds of sound.}

In addition, for the $\cssq=10^{-4}$ case, the results from \code{CLASS} and \kev are indistinguishable at least until $z=0.5$, and start to differ only slightly at $z=0$ {at large scales}. 
For the $\cssq=10^{-7}$ case, while the results from \code{CLASS} and \kev are consistent at both high- and low-$k$, there is a different transition from $\mu>1$ to $\mu=1$ in \kev compared to \texttt{CLASS}. 
{This is because for the lower sound speed the sound-horizon lies within the scale of matter non-linearity. As dark matter clustering becomes non-linear, $\delta_m$ becomes much larger than in linear theory, which reduces $\Phi/\delta_m$ and hence $\mu$. This can be mimicked by turning on Halofit, and indeed using \code{CLASS} with Halofit to extract $\mu$ allows to match the result of \kev better at scales around $k\approx0.1 h/$Mpc, as we can see in panel (d) of Fig.~\ref{figmu}.}

{On even smaller scales, $k\gtrsim 1h/$Mpc, the combination of \code{CLASS} with Halofit undershoots the \kev result. This is due to non-linear clustering of the $k$-essence field on small scales. This can only be correctly modeled by simulating the $k$-essence field itself. Also {\it gevolution} 1.2 with the \code{CLASS} interface, where $k$-essence is a realisation of the linear spectrum, is not able to simulate this region correctly.}

\subsection{A fitting function for $\mu$ in \kess}

To characterize the contribution of $k$-essence to the gravitational potential, and to simplify the inclusion of non-linear $k$-essence clustering in linear Boltzmann codes and in back-scaled Newtonian $N$-body simulations (see next section), we approximate the numerical $\mu(k,z)$ with a simple fitting formula.
The fact that we recover GR/$\Lambda$CDM at small scales and that we have a constant modification to GR at large scales motivates us to choose a function that smoothly connects two different regimes, and we propose the following fitting function for $\mu(k,a)$:
\be
f(k|\alpha,\beta,\gamma) = 1+\alpha \left(1-\tanh \big(\beta (\log_{10} k-\gamma) \big)\right) \, .
\label{eq:muparam}
\ee
Here $\alpha$ controls the amplitude of $\mu$ on large scales, $\gamma = \log_{10}\kappa$ the location of the transition, and $\beta$ the steepness of the transition.
The variables $\alpha$, $\beta$, and $\gamma$ depend on time as well as on the cosmology. We can use the $(\alpha,\beta,\gamma)$ parameter space to distinguish between models / constrain cosmology. {Additionally, the function (\ref{eq:muparam}) is $C^\infty$ and its derivatives are easily computed, while numerical derivatives of simulation results are often noisy.} In Appendix \ref{fitting_func} we discuss how these parameters control the shape of $\mu$, and how well the fitting function describes the simulation results. {We find that  for $k$-evolution and \code{CLASS}, the parametrisation (\ref{eq:muparam}) is able to describe $\mu(k,z)$ at the sub-percent level relative to $\mu$. }

\begin{figure}[tb]%
    \centering
   {{\includegraphics[scale=0.36]{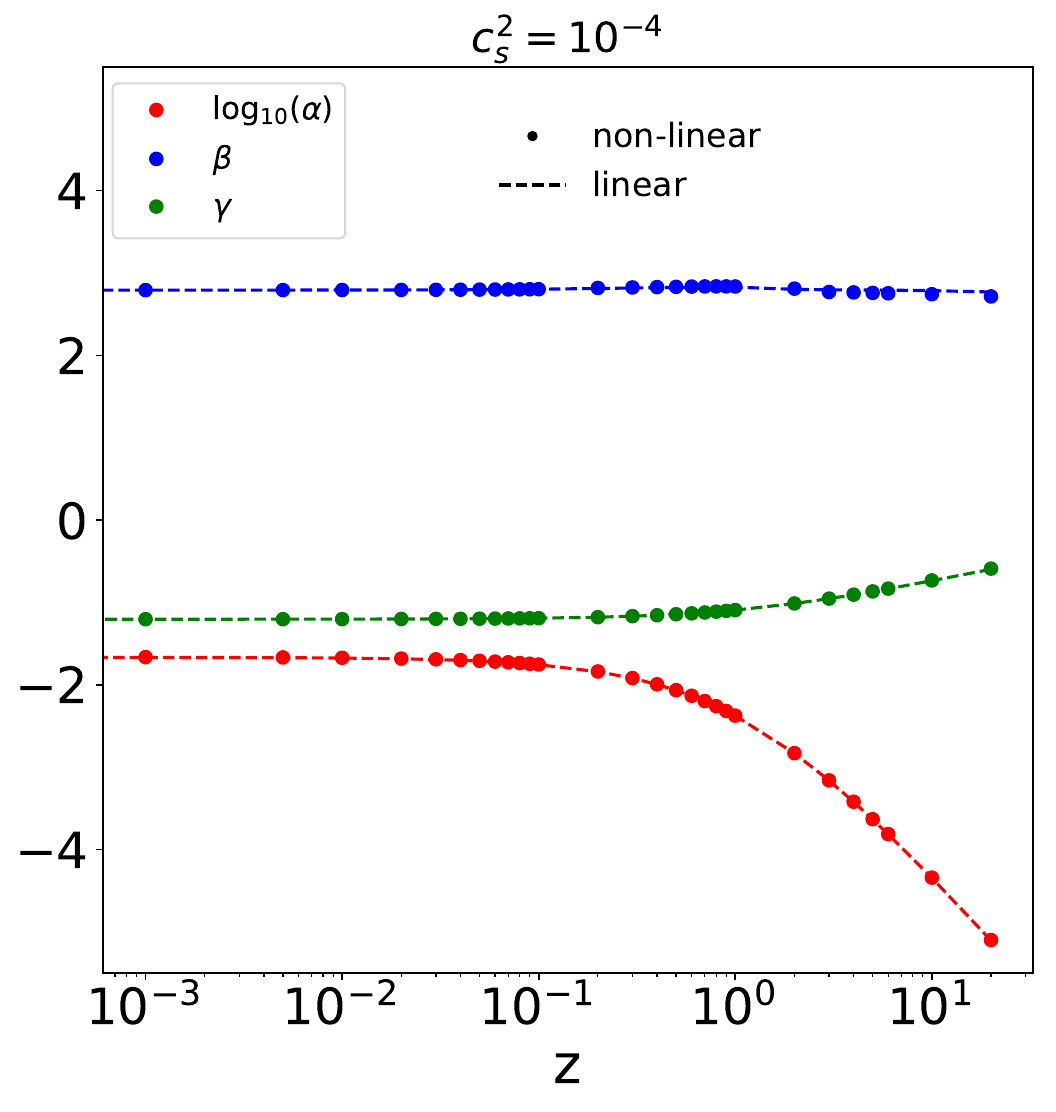} }}%
    \qquad
   {{\includegraphics[scale=0.36]{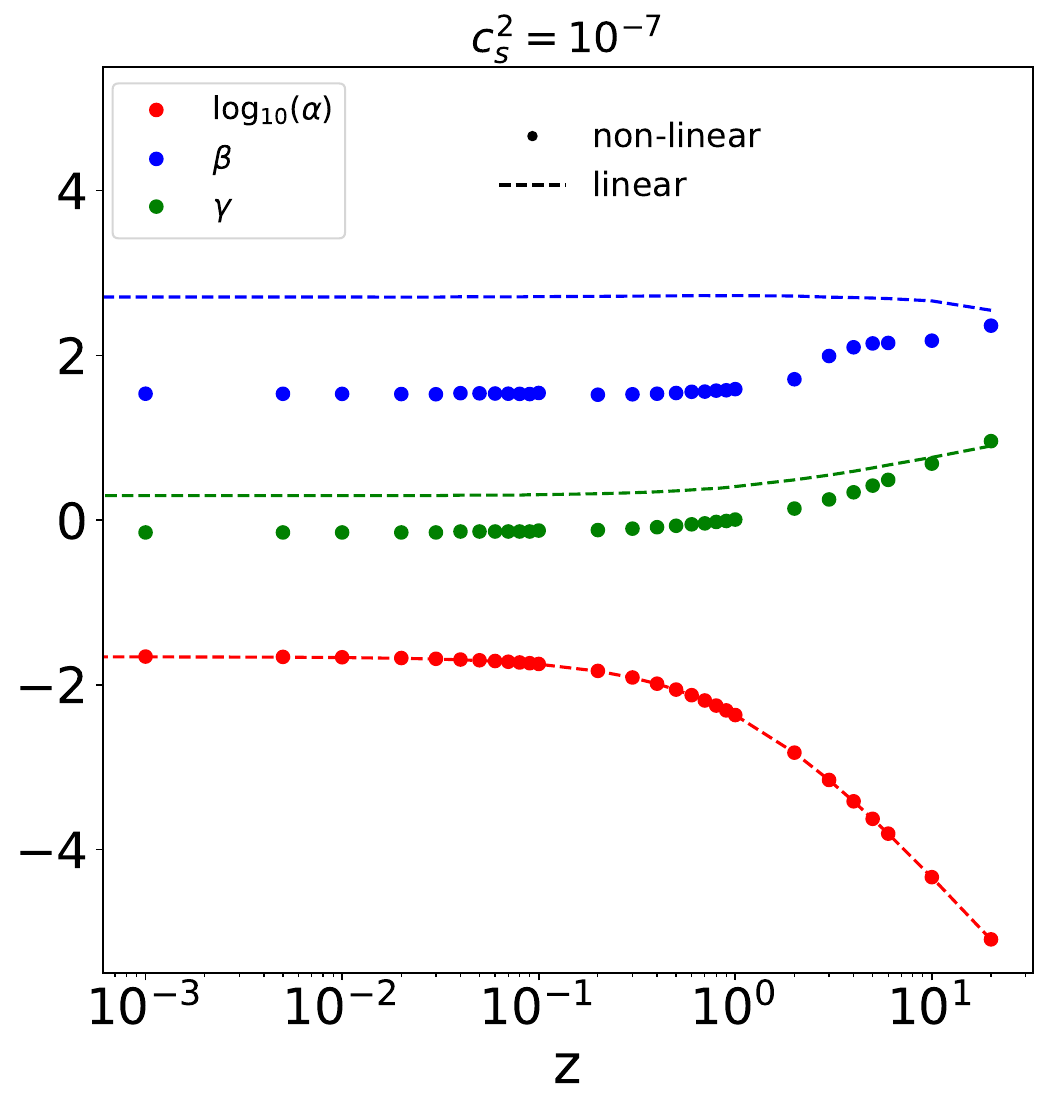} }}%
        \caption{The fitted parameter values for the $\mu$ parametrization (\ref{eq:muparam}), for the two $k$-essence models with  $c_s^2=10^{-4}$ (left) and $c_s^2=10^{-7}$ (right), as a function of redshift. 
        Points and dashed lines are respectively the results of $k$-evolution and \code{CLASS}. }
    \label{fitting_mu}%
\end{figure}

The fit enables us to model $\mu(k,a)$ in a simple way and to describe its evolution by studying the time and scale evolution of the parameters.
We perform the fit in the non-linear ($k$-evolution) and linear (\code{CLASS}) cases. 
Fig.~\ref{fitting_mu} shows the evolution of three fitting parameters as a function of redshift for both the linear (solid lines) and non-linear (dots) cases.
As expected from Fig.~\ref{figmu}, for $c_s^2=10^{-4}$, there is little difference between the linear and non-linear cases.  
Also for the $c_s^2=10^{-7}$ case, the fitted amplitudes ($\alpha$) are consistent between the linear and non-linear cases. 
Most of the difference arises in the steepness ($\beta$) and the location ($\gamma$) of the transition. 
In  Appendix \ref{fitting_func} we provide an additional figure, Fig.~\ref{fig:mu_params_evolution} that shows in more details the evolution of the parameters. In that figure we can see that there are also detectable differences in $\beta$ between the linear and \kev results for $c_s^2=10^{-4}$, but that they are relatively small. In Table.~\ref{table_param} we show the parameters ($\alpha,\beta,\gamma$) fitted to the $k$-evolution data at redshifts $z=0$ and $z=1$ for both speeds of sound. The full redshift information for the fitting parameters are delivered as a text file in the arXiv submission.

\noindent
\begin{table}
\begin{center}
\resizebox{10cm}{!} {
\centering
\begin{tabular}{cc|c|c|c|l}
\cline{3-5}
& & \multicolumn{3}{ c| }{Parameters} \\ \cline{3-5}
& & $\alpha$ & $\beta$ & $\gamma$ \\ \cline{1-5}
\multicolumn{1}{ |c  }{\multirow{2 }{*}{ $c_s^2 =10^{-4}$} } &
\multicolumn{1}{ |c| } {\bluet $z=0$} & \bluet $0.021$\quad &  \bluet $2.79$ & \bluet $-1.20$ &      \\ \cline{2-5}
\multicolumn{1}{ |c  }{}                        &
\multicolumn{1}{ |c| }{  \redt $z=1$} & \redt  $0.00422$ & \redt $2.83$ & \redt $-1.10$ &      \\ \cline{1-5}
\\\cline{1-5}
\multicolumn{1}{ |c  }  { \multirow{2}{*}{$c_s^2 =10^{-7}$} } &
\multicolumn{1}{ |c| }{ \bluet $z=0$} & \bluet $0.020$ & \bluet  $1.46$ &  \bluet $-0.154$ &  \\ \cline{2-5}
\multicolumn{1}{ |c  }{}                        &
\multicolumn{1}{ |c| }{ \redt $z=1$} & \redt $0.00429$ & \redt $1.53$ & \redt $0.0069$ &   \\ \cline{1-5}
\end{tabular}
}
\end{center}
\caption{Parameter values fitted to \kev results for both speeds of sound at two redshifts, $z=0$ and $z=1$.}
\label{table_param}
\end{table}
\section{Applications of $\mu(k,z)$\label{sec:use}}
In this section we discuss how one can use the $\mu (k,z)$ parametrisation in combination with other codes, especially linear Boltzmann and Newtonian $N$-body codes. To answer this question, we first illustrate the differences between these codes in Fig.~\ref{picture_Poissoneq} when there is \kess as a dark energy candidate. In $k$-evolution all the components including short-wave corrections, relativistic terms, matter and \kess non-linearities are included. In Newtonian $N$-body codes, the equations are solved in N-body gauge, see Appendix \ref{N-body_gauge} for more details. In these codes one can capture the background evolution correctly, while short wave-corrections are absent and $k$-essence perturbations are at most taken into account through the initial conditions.
In the linear Boltzmann codes, on the other hand, non-linearities in matter and \kess as well as short wave corrections are absent.
\begin{figure}[tb]%
    \centering
    \hspace*{-1cm}  
   {{\includegraphics[scale=0.29]{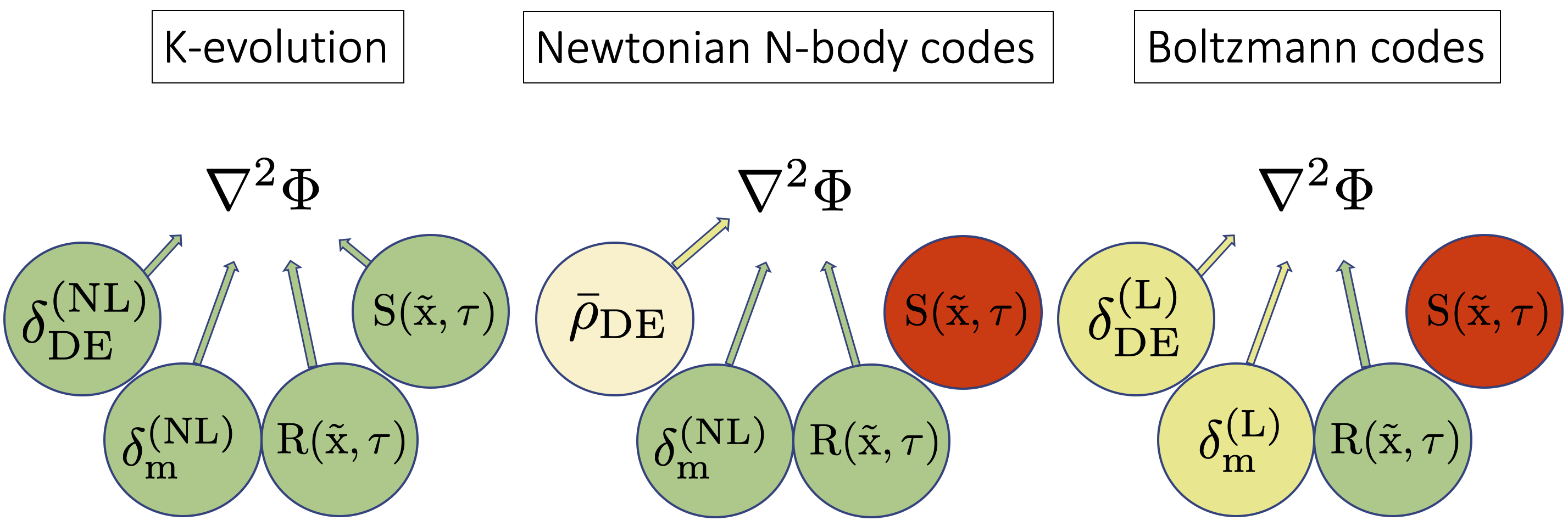} }}%
        \caption{This picture compares the way that different codes solve the Poisson equation. On the left we have the relativistic $N$-body code \kev which uses the full Poisson equation including non-linear $k$-essence and matter densities, relativistic terms and short wave corrections (all in green). Standard Newtonian $N$-body codes (middle) solve the Poisson equation for the correct background expansion rate (term in light yellow), and include non-linear matter densities as well as relativistic terms in N-body gauge (terms in green), but do not take into account $k$-essence perturbations or shortwave (nonlinear GR) terms (in red). Linear Boltzmann codes like \class (on the right) are fully relativistic and include linear density perturbations for both matter and $k$-essence (in yellow), but no non-linear GR and matter contributions (in red). }
    \label{picture_Poissoneq}%
\end{figure}

\subsection{Improving linear Boltzmann codes and Fisher forecasts with a parametrised $\mu$}

Recipes to predict the non-linear matter power spectrum are routinely implemented in Boltzmann codes in order to source weak lensing calculations, which are themselves linear but very sensitive to small-scale power. The $\mu(k,z)$ function presented in the previous section can then be used to correct for the non-linear effect of $k$-essence on the lensing. For example, $\mu(k,z)$ would appear as a simple factor in the line-of-sight integral for the lensing potential if the non-linear matter power spectrum is already calibrated for the correct background model. The same correction can also be applied in the context of Fisher forecasts.

\subsection{Improving Newtonian simulations}

In the companion paper \cite{kevolution} we have shown that the so-called backscaling method to set initial conditions in Newtonian $N$-body codes is able to recover the correct non-linear matter power spectrum in $k$-essence models.
As explained in detail in \cite{kevolution},
this is achieved by scaling back the linear power spectrum at the final redshift using the linear growth function in the given background model. While this works well for the matter power spectrum, it is impossible to simultaneously obtain an accurate power spectrum of the gravitational potential, as the latter is sourced additionally by $k$-essence perturbations. Our $\mu$ precisely parametrises the correction necessary to obtain the potential from the matter power spectrum, and additionally allows to reconstruct the power spectrum of $k$-essence perturbations. An immediate practical application would be to include this correction when calibrating emulators like \cite{Knabenhans:2018cng}.

\section{Conclusions}
In this paper we have studied metric perturbations in the weak-field regime of General Relativity, in the presence of a $k$-essence scalar field as dark energy. We showed that the short-wave corrections to the Hamiltonian constraint are negligible at all redshifts and all scales, while the relativistic terms are only relevant at large scales, leaving the terms with Poisson-gauge matter and \kess density perturbations as the main source at quasi-linear and small scales. The relativistic terms and the density perturbations can be combined, in the usual way, to form a linear Poisson equation that then holds on all scales of interest in cosmology.

We study the contribution of the $k$-essence scalar field to the metric perturbations through the $\mu$ parametrisation that encodes the additional contribution of a dark energy fluid or a modification of gravity to the Poisson equation. As the Poisson equation is valid on all scales, this description works even at non-linear scales if $\mu$ is understood as an average effect.
We show that for $k$-essence fields with a high speed of sound, the linear theory agrees with simulation results, while for models with a small speed of sound we see large deviations from linear theory. Our results are thus important for tests of low speed of sound $k$-essence models with future surveys.

We encode our $k$-essence simulation results for $\mu(k,z)$ in an easy-to-use $\tanh$-based fitting function, together with recipes on how to include the function in linear Boltzmann codes or Newtonian $N$-body simulations with different expansion rate but without additional $k$-essence field. 
{While in this paper we only consider two $k$-essence models with different speeds of sound, we plan to provide fits to $\mu(k,z)$ for a wider range of models in a follow-up publication.}
\setcounter{equation}{0}
\section*{Acknowledgements}
FH would like to thank Jean-Pierre Eckmann for helpful comments about manuscript and useful discussions. FH and MK acknowledge financial support from the Swiss National Science Foundation. This work was supported by a grant from the Swiss National Supercomputing Centre (CSCS) under project ID s710. BL would like to acknowledge the support of the National Research Foundation of Korea (NRF-2019R1I1A1A01063740). AS would like
to acknowledge the support of the Korea Institute for Advanced Study (KIAS) grant funded by the Korean government. {JA acknowledges funding by STFC Consolidated Grant ST/P000592/1.}

 \appendix

\section{Discussion about N-body gauge} \label{N-body_gauge}
The correspondence between Newtonian $N$-body simulations and GR can be established through a particular gauge, called the N-body gauge \cite{Fidler:2016tir}, in which, generally speaking, one requires
\be
\label{eq:NbPoisson}
    \nabla^2 \Phi_N = 4 \pi G_N \bar{\rho}_m \delta_m^\mathrm{count}\,,
\ee
and
\be
    \frac{dv^i}{d\tau} + \mathcal{H} v^i = -\nabla \Psi\,,
\ee
where in the first equation $\Phi_N$ is the contribution of non relativistic matter to $\Phi$, and $\delta_m^\mathrm{count}$ is a counting density (rest mass per coordinate volume). With two scalar gauge generators $L$ and $T$ at one's disposal, where $\tau \rightarrow \tau + T$ and $x^i \rightarrow x^i + \nabla^i L$ are the coordinate transformations, it turns out that under quite generic conditions there is a family of gauge transformations that satisfy the above two conditions at the linear level. 

To illustrate this, let us start from the Poisson gauge and find $T$, $L$ such that above equations hold. The first thing to note is that $\Psi$ corresponds to a gauge-invariant variable (we are working at linear order), so the second equation already holds in Poisson gauge. Since velocities transform as $v^i \rightarrow v^i + \nabla^i L'$ maintaining the form of the second equation readily requires $L' \simeq 0$.

Before turning to the first equation, let us define the scalar metric perturbations of the N-body gauge as follows:
\be
    ds^2 = a^2 \left[-\left(1 + 2 A^\mathrm{Nb}\right) d\tau^2 -2 \nabla_i B^\mathrm{Nb} dx^i d\tau + \left(1 + 2 H_L^\mathrm{Nb}\right) \delta_{ij} dx^i dx^j - 2 \left(\nabla_i \nabla_j - \frac{\delta_{ij}}{3} \nabla^2\right)H_T^\mathrm{Nb} dx^i dx^j\right]
\ee
Noting how the various scalar perturbations transform, for a $T$ and $L$ connecting the N-body gauge to Poisson gauge we have
\be
    \Phi = -H_L^\mathrm{Nb} - \mathcal{H} T - \frac{1}{3} \nabla^2 L\,,
\ee
\be
    \Psi = A^\mathrm{Nb} + \mathcal{H} T + T'\,,
\ee
\be
    0 = B^\mathrm{Nb} + T - L' \Rightarrow B^\mathrm{Nb} \simeq  -T\,,
\ee
\be
    0 = H_T^\mathrm{Nb} - L \Rightarrow H_T^\mathrm{Nb} = L\,,
\ee
and
\be
    \delta_X = \delta_X^\mathrm{Nb} - 3 \left(1 + w_X\right) \mathcal{H} T\,.
\ee
We can insert these expressions into the Hamiltonian constraint (keeping the term $\nabla^2\Phi$ in place) to obtain
\be
    \nabla^2 \Phi + 3 \mathcal{H} \left({H_L^\mathrm{Nb}}' + \mathcal{H} T' + \mathcal{H}' T\right)
    - 3 \mathcal{H}^2 \left(A^\mathrm{Nb} + \mathcal{H}T + T'\right) = 4 \pi G_N a^2 \sum_X \bar{\rho}_X\left[\delta_X^\mathrm{Nb} - 3 \left(1 + w_X\right) \mathcal{H} T\right]\,,
\ee
where we already used the condition $L' \simeq 0$. Noting that
\be
\mathcal{H}^2 - \mathcal{H}' = 4 \pi G_N a^2 \sum_X \bar{\rho}_X \left(1+w_X\right)
\ee
we immediately get
\be
    \nabla^2 \Phi + 3 \mathcal{H} {H_L^\mathrm{Nb}}' - 3 \mathcal{H}^2 A^\mathrm{Nb} = 4 \pi G_N a^2 \sum_X \bar{\rho}_X \delta_X^\mathrm{Nb} = 4 \pi G_N a^2 \bar{\rho}_m \left(\delta_m^\mathrm{count} - 3 H_L^\mathrm{Nb}\right) + 4 \pi G_N a^2 \sum_{X\neq m} \bar{\rho}_X \delta_X^\mathrm{Nb}
\ee
The requirement that this equation is compatible with eq.~(\ref{eq:NbPoisson}) does not yet fix the gauge completely. One may try to require that $H_L^\mathrm{Nb} \simeq 0$ which means that $\delta_m^\mathrm{Nb} = \delta_m^\mathrm{count}$. This means that not only are the Newtonian equations satisfied, but also the counting density \textit{is} the physical density in that gauge.

One can easily see from the last equation that $H_L^\mathrm{Nb} \simeq 0$ also suggests $A^\mathrm{Nb} \simeq 0$, and one can verify that, as long as pressure perturbations are small, this condition can be met \cite{Fidler:2016tir}. We can then infer $T$ and $L$ as follows.
\be
 \Psi = \mathcal{H} T + T'\,, \qquad \Phi = -\mathcal{H} T - \frac{1}{3} \nabla^2 L\,,
\ee
hence
\be
 \mathcal{H} \Psi + \Phi' = \left(\mathcal{H}^2 -\mathcal{H}'\right) T\,,
\ee
where $L' \simeq 0$ was assumed. We can now see that the momentum constraint implies $T = -\nabla^{-2} \theta_\mathrm{tot}$ and hence $\delta_m^\mathrm{Nb} = \Delta_m$. Inserting $T$ back into its relation with $\Phi$ above, we also get
\be
    \nabla^2 L = 3 \mathcal{H} \nabla^{-2} \theta_\mathrm{tot} - 3 \Phi\,.
\ee
The right-hand side is the comoving curvature perturbation which is indeed conserved at late times in standard cosmology. Hence our assumption $L' \simeq 0$ was consistent.
} 

\section{Details of the fitting function for $\mu(k,z)$ }
\label{fitting_func}

In this Appendix we discuss the properties of $\tanh$ fitting function, we also compare the fitted values to the underlying simulation results and show that the $\tanh$ fitting function works quite well. As explained in more detail in the main text, we need a function to smoothly connect two different regimes, namely, between $\mu\simeq$ constant on large scales to $\mu=1$ on small scales. To do this, we propose the following form, at a fixed redshift:
\be
f(k|\alpha,\beta,\gamma) = 1+\alpha \left(1-\tanh \big(\beta (\log_{10} k-\gamma) \big)\right).
\ee
The parameter $\gamma = \log_{10}\kappa$ determines the location of the transition, $\beta$ the steepness of the transition, while $1+2 \alpha$ is the value of $\mu$ for $k\rightarrow 0$. All of these parameters are functions of redshift (or scale factor).
The fit enables us to describe the scale dependence of $\mu(k,a)$ in a simple way, while the time dependence can be studied through the evolution of fit parameters with redshift. 

\begin{figure}[H]
    \centering
    {{\includegraphics[scale=0.33]{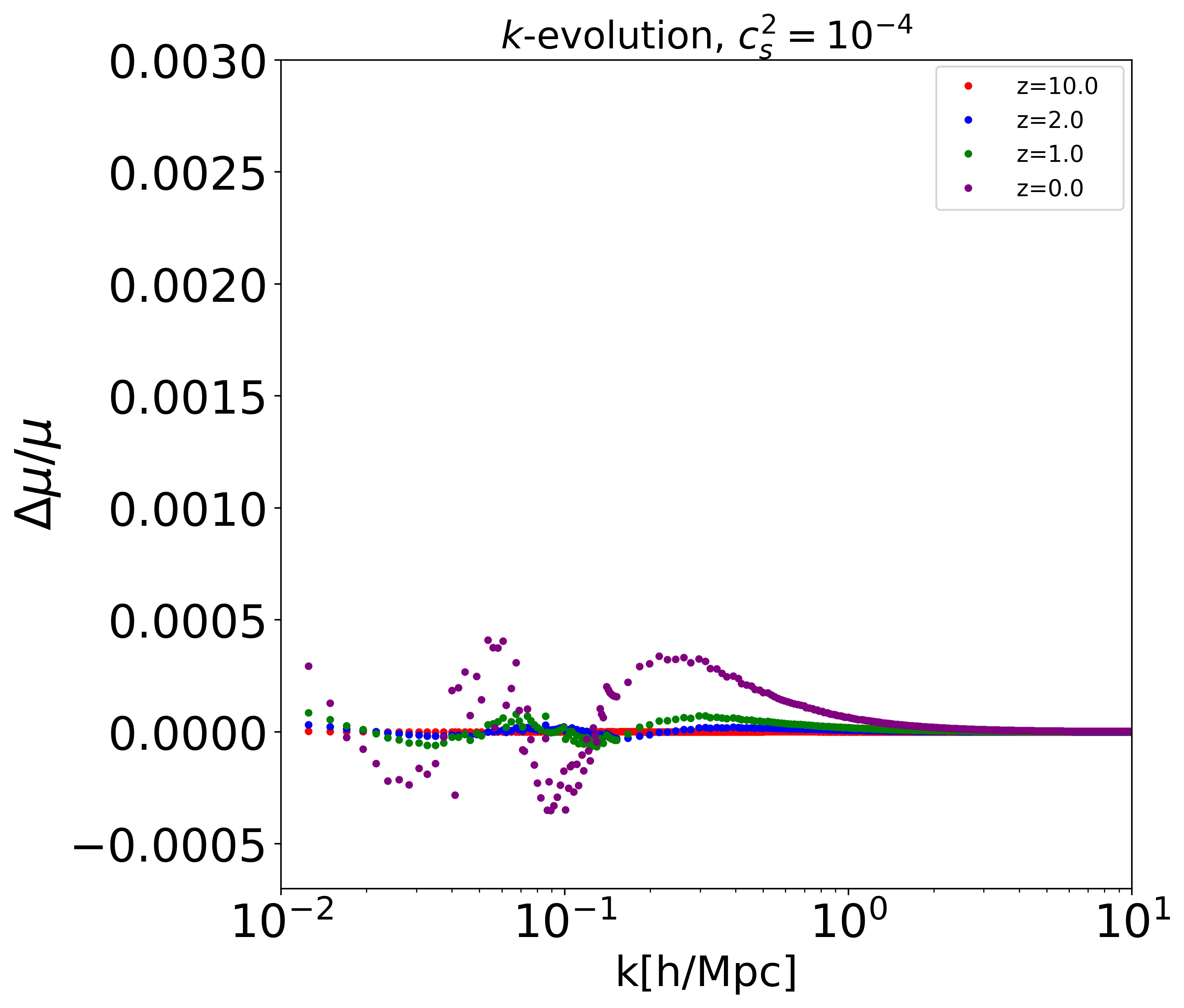} }}%
    \qquad
    {{\includegraphics[scale=0.33]{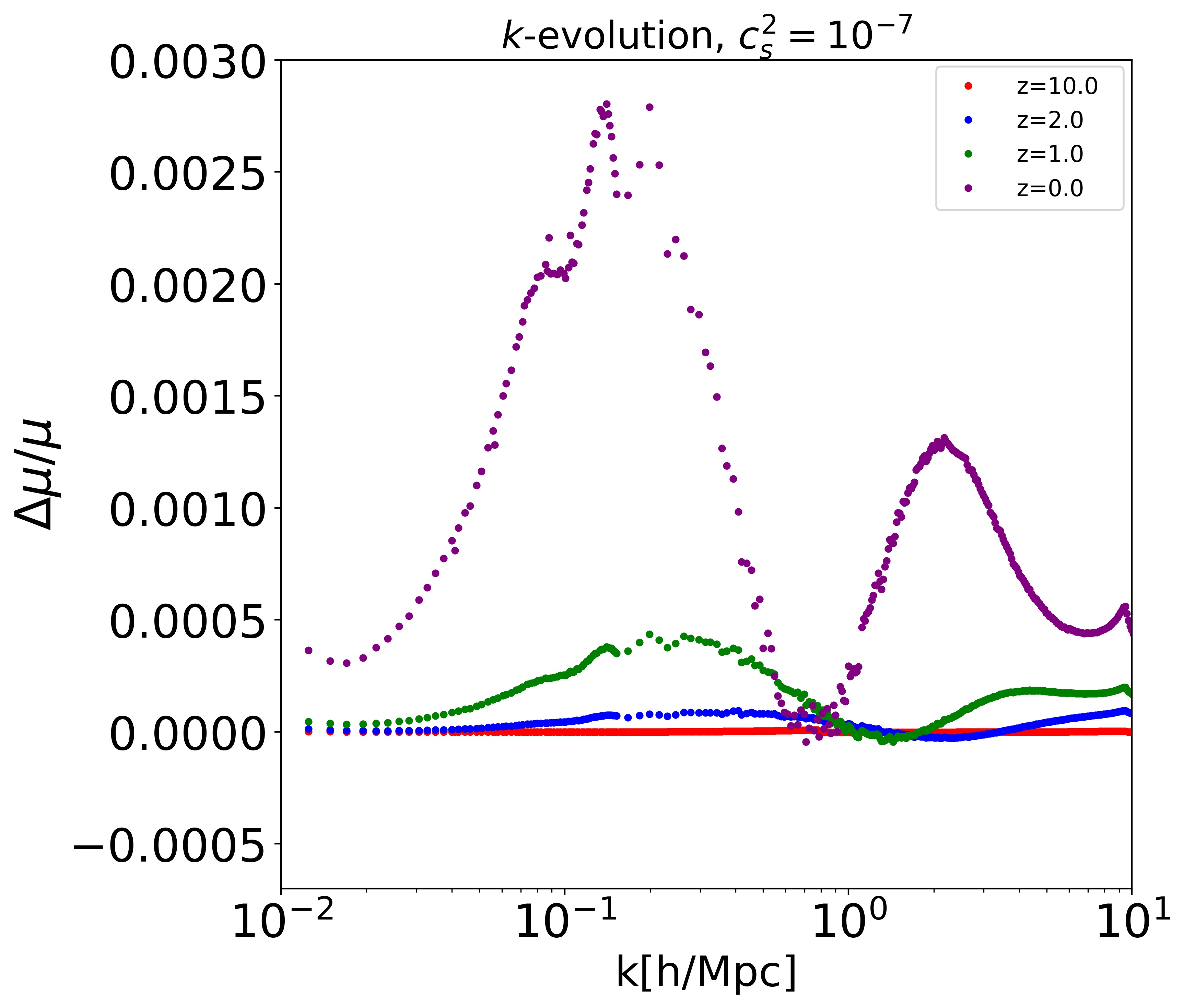} }}%
        \caption{The relative error for the fits compared to the actual $\mu$ obtained from $k$-evolution, for two different speeds of sound, $c_s^2 = 10^{-7}$ (right) and $c_s^2 = 10^{-4}$ (left). {The accuracy of the fit is of the sub-percent level for both speeds of sound at all redshifts and all scales.}}
    \label{fitting_mu_validity1}%
\end{figure}

\begin{figure}[H]
    \centering
    {{\includegraphics[scale=0.33]{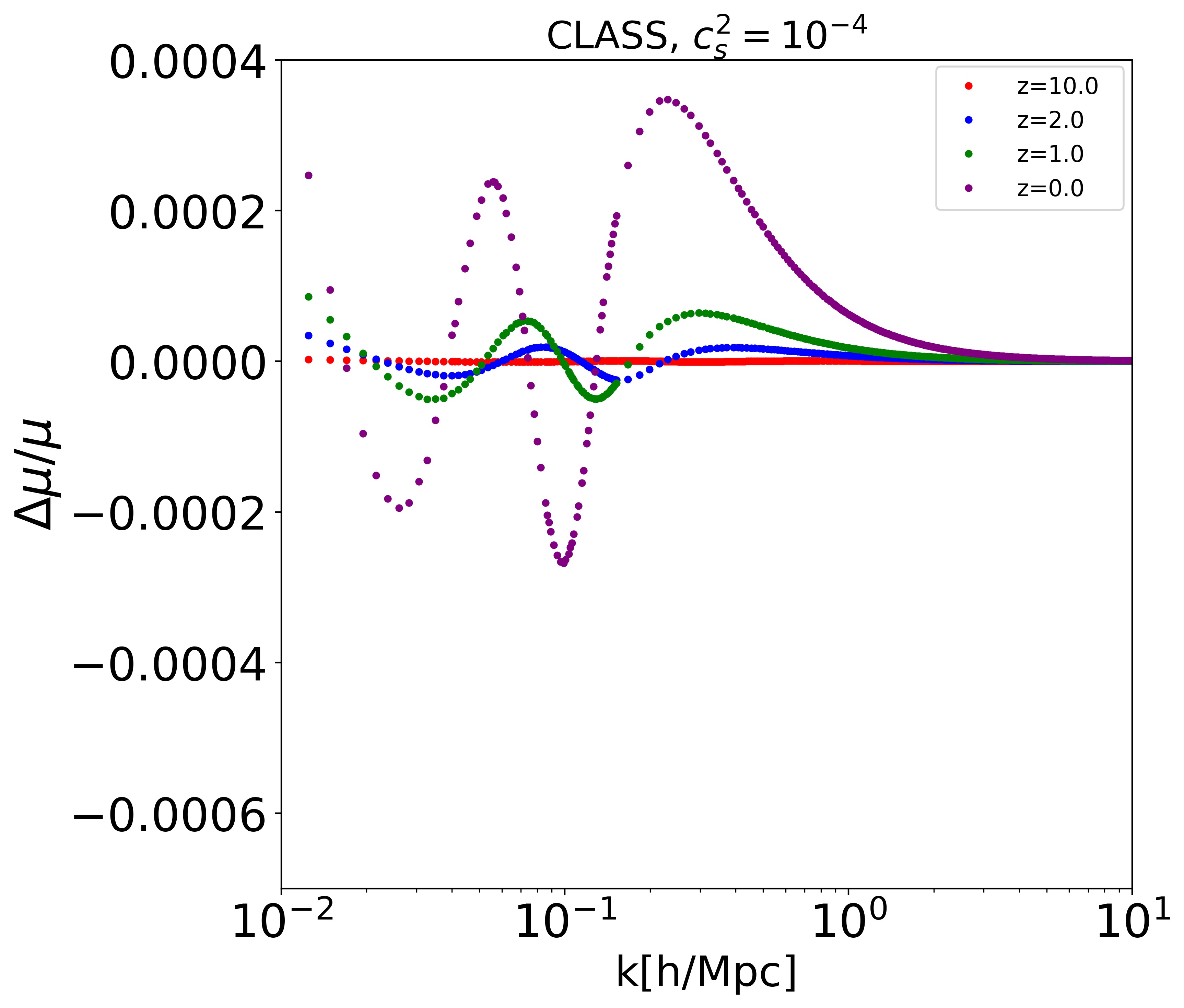} }}%
    \qquad
    {{\includegraphics[scale=0.33]{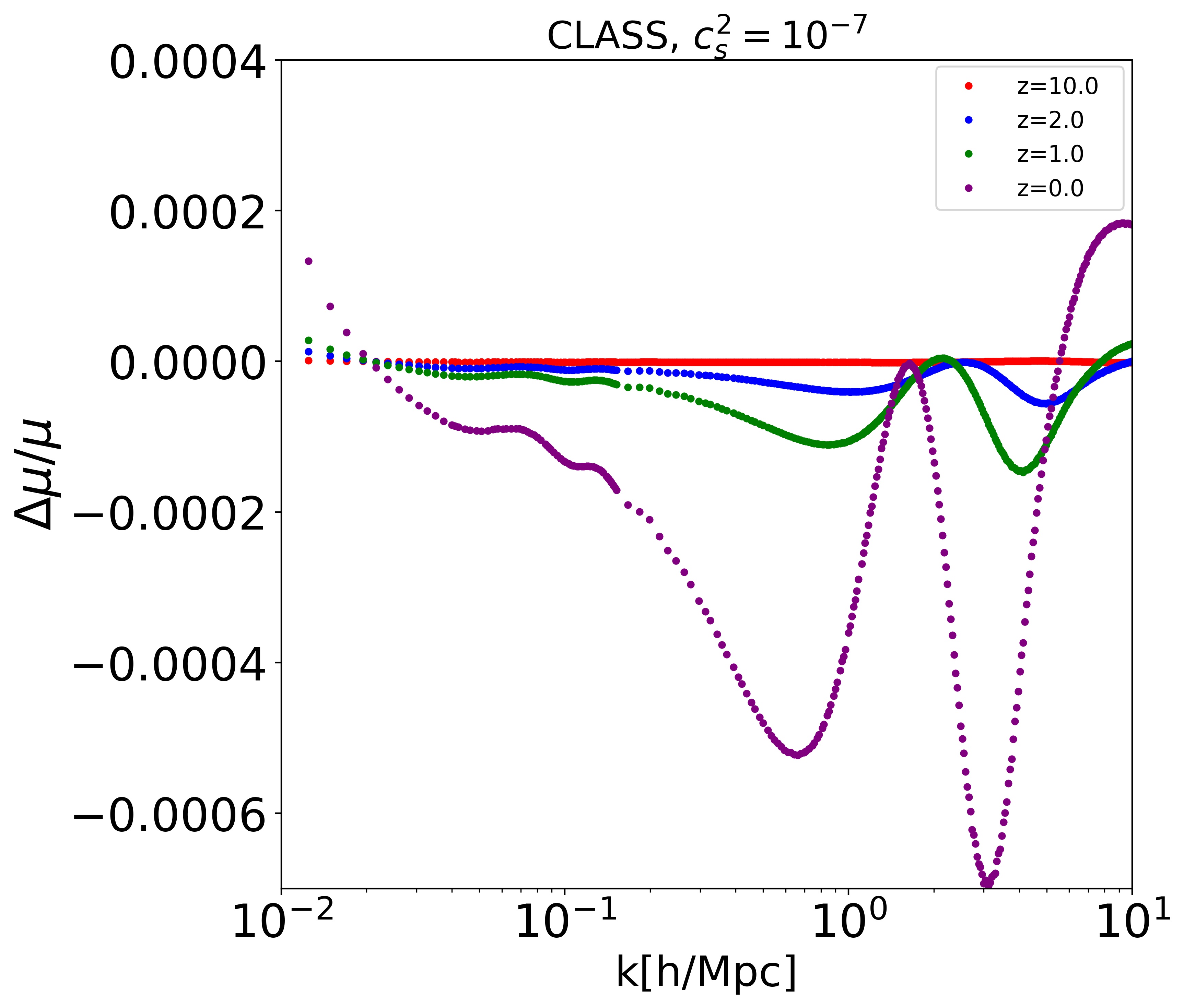} }}%
        \caption{The relative error of the fits compared to \code{CLASS} results for two speeds of sound, $c_s^2 = 10^{-7}$ (right) and $c_s^2 = 10^{-4}$ (left).  {The results show that the accuracy of the fit is of the sub-percent level for both speeds of sound at all redshifts and all scales.} }
    \label{fitting_mu_validity2}%
\end{figure}

\begin{figure}[h!]
    {{\includegraphics[scale=0.11]{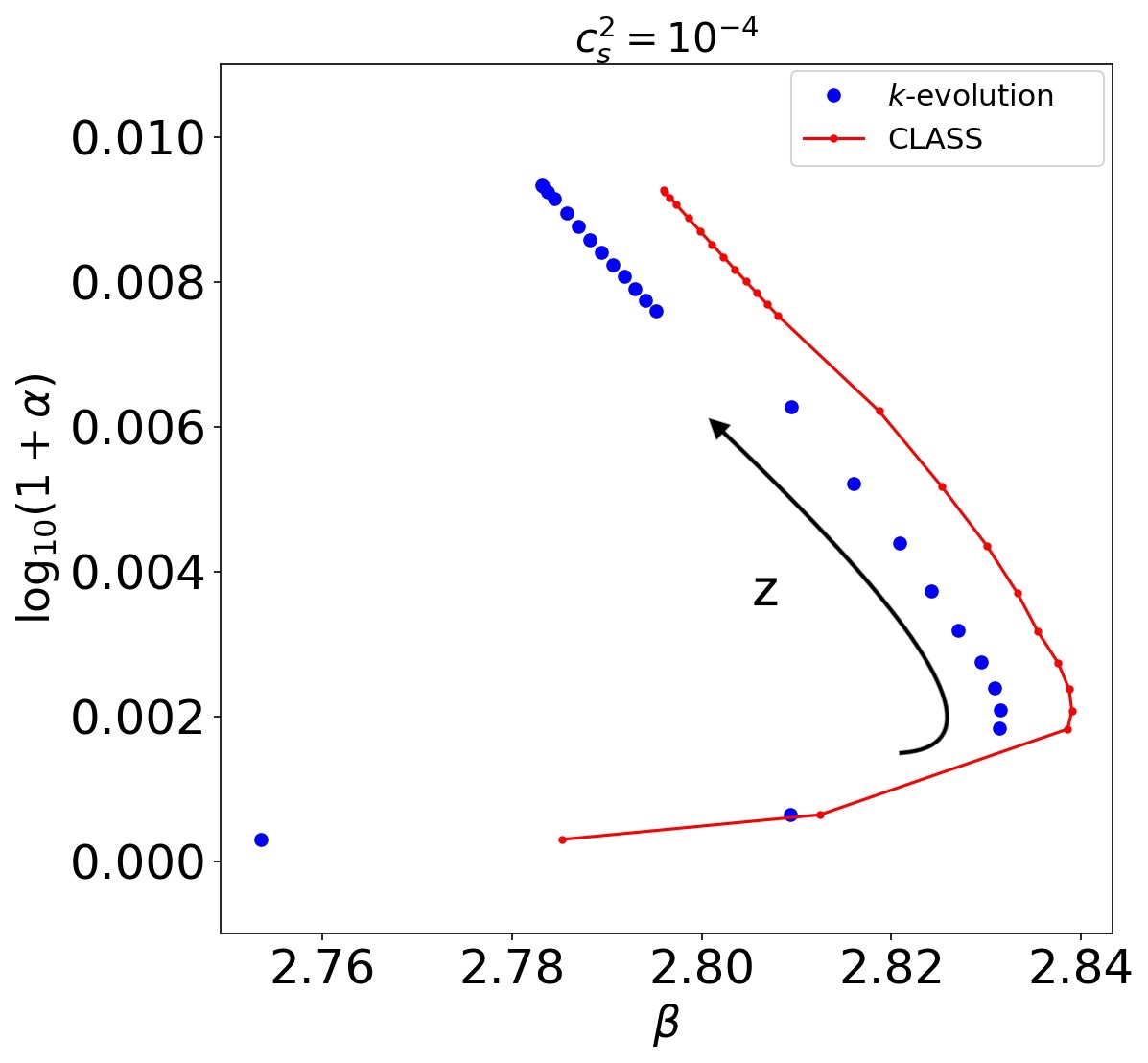}}}%
        \qquad
    {{\includegraphics[scale=0.11]{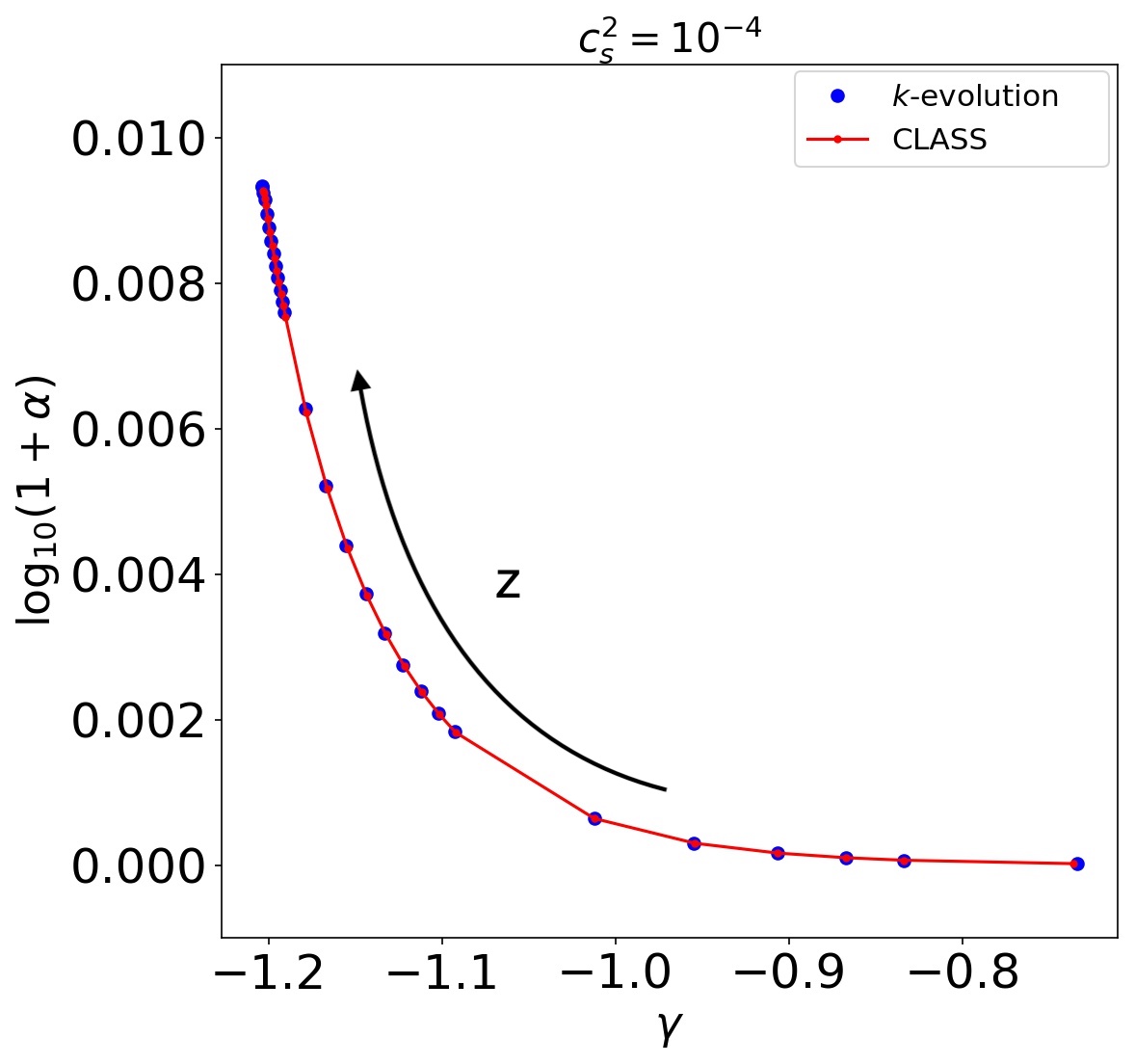}}}%
        \qquad    
    {{\includegraphics[scale=0.11]{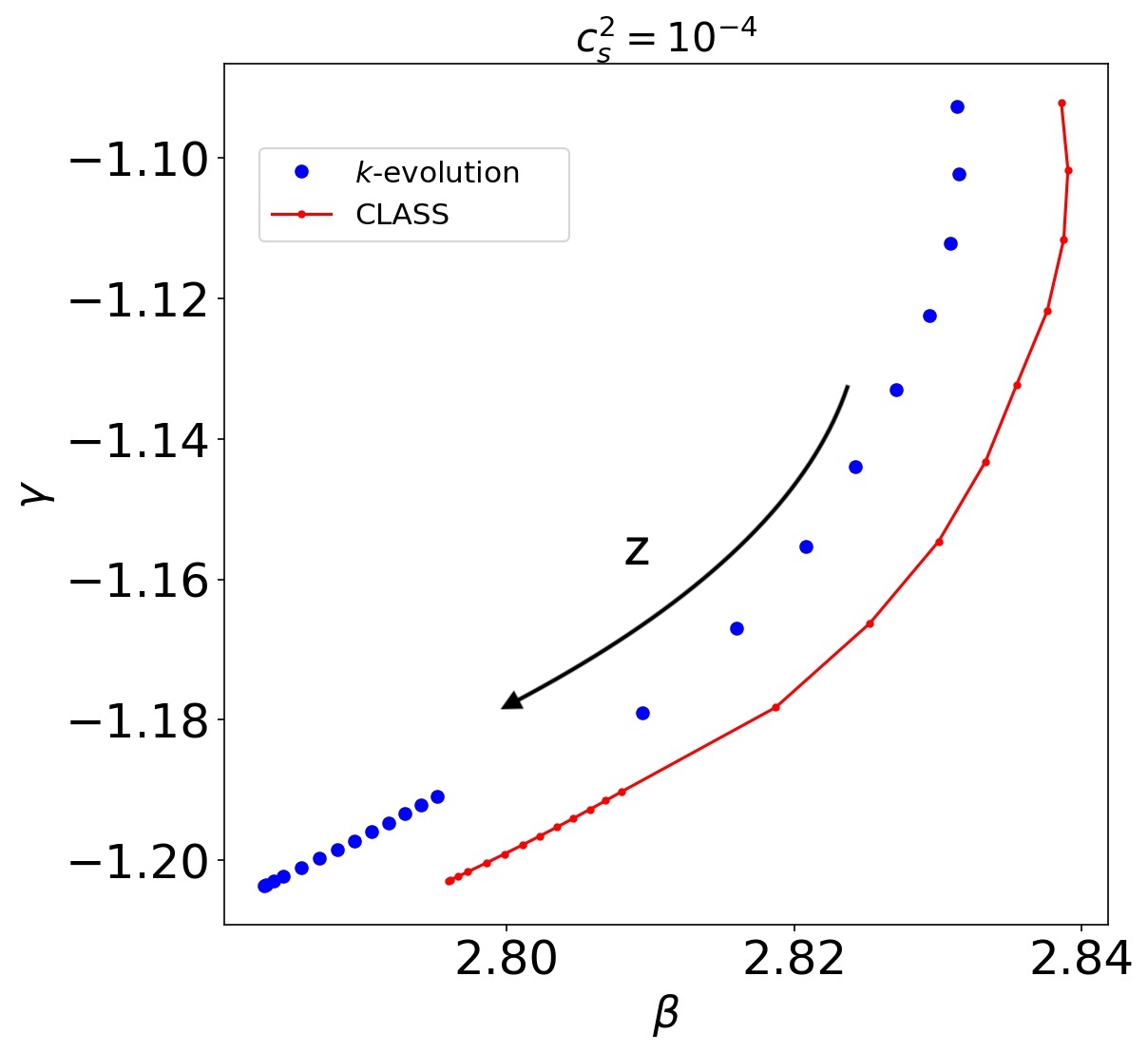}}}%
       \qquad
    {{\includegraphics[scale=0.11]{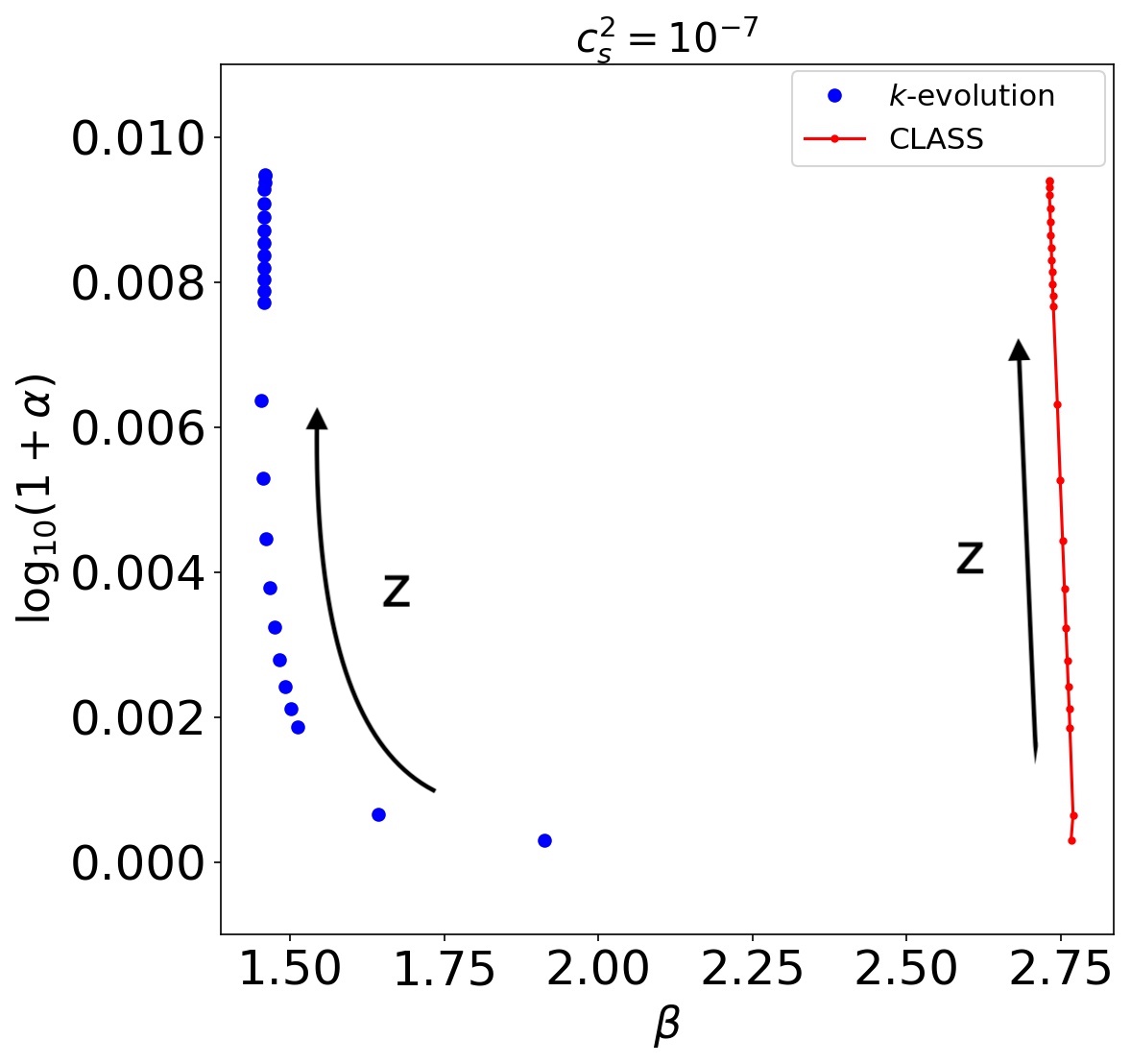} }}%
    \qquad
    {{\includegraphics[scale=0.11]{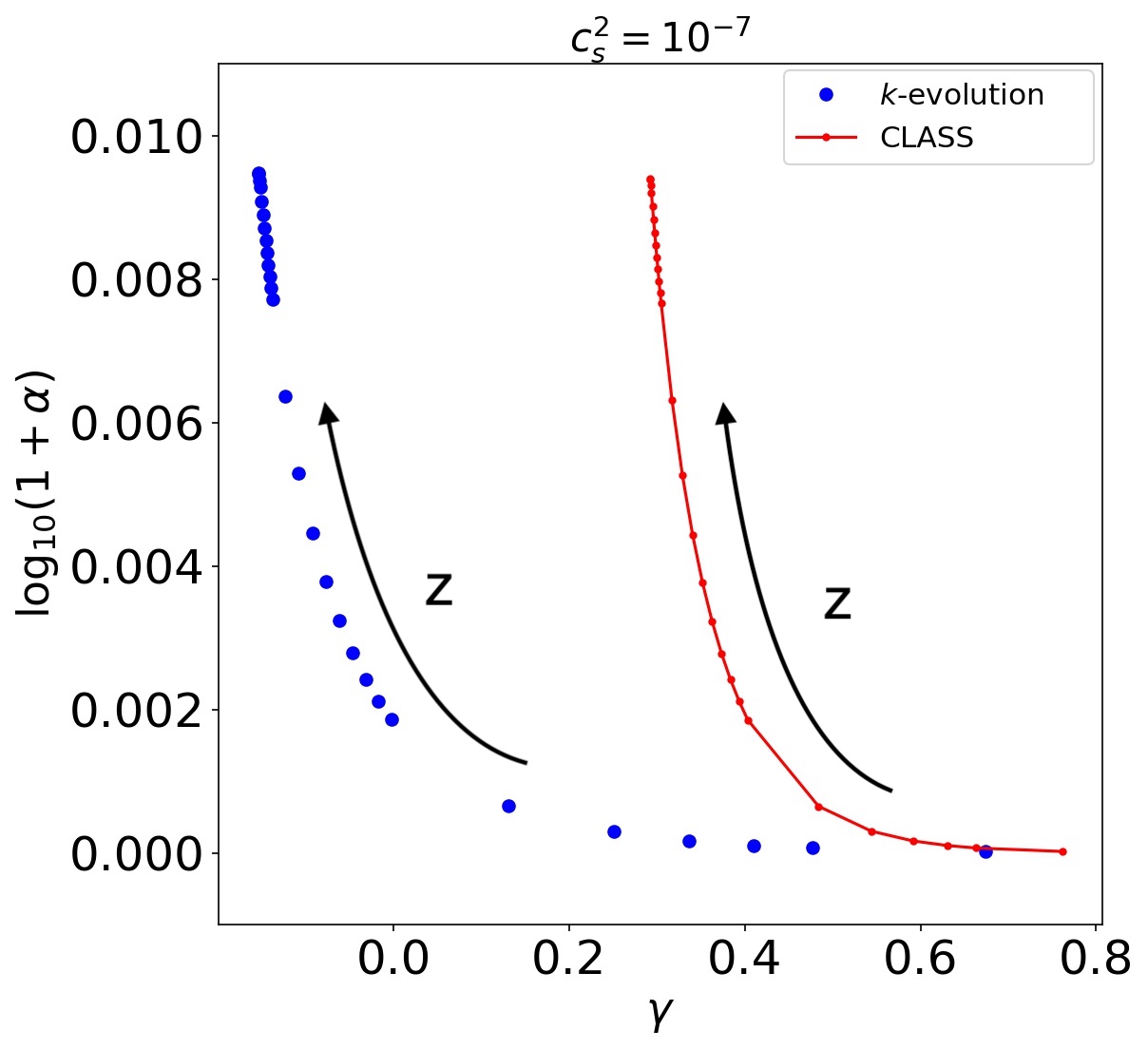} }}%
    \qquad    
    {{\includegraphics[scale=0.11]{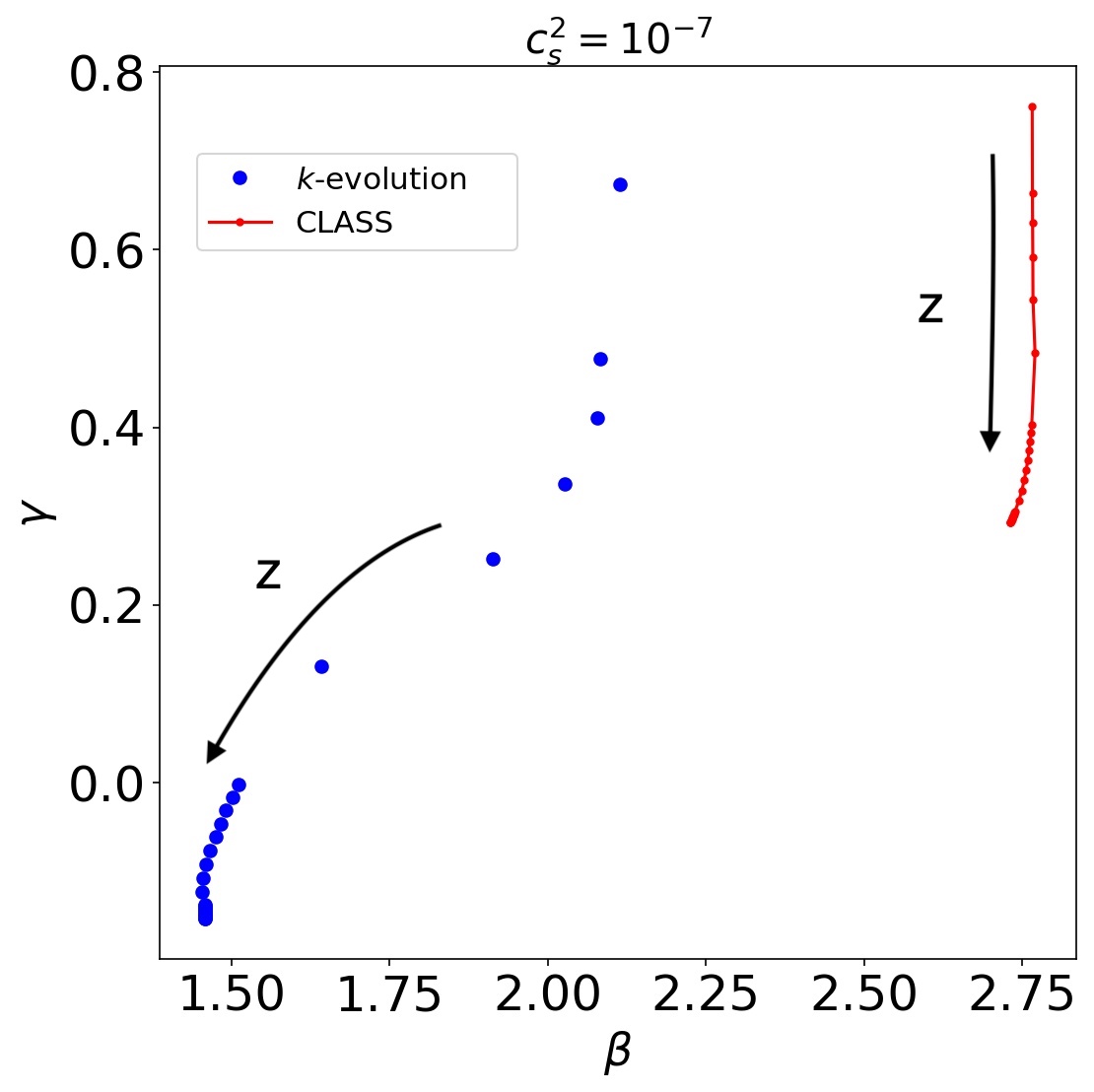}}}%
        \caption{ Evolution of the fit parameters $\log_{10}(1 + \alpha)$, $\beta$, and $\gamma$ for $k$-evolution and \code{CLASS} data for two speeds of sound $c_s^2=10^{-4}$ (top) and $10^{-7}$ (bottom) is shown. The arrows show the direction of decreasing redshift (increasing time). For the case of low speed of sound {(bottom panel)} the parameter values obtained from \kev and from the linear code \class differ, while for large speed of sound {(top panel)} the parameters almost match, suggesting that one could simply use linear Boltzmann codes for treating high speed of sound $k$-essence models.
        }
    \label{fig:mu_params_evolution}%
    
\end{figure}

In Fig.~\ref{fitting_mu_validity1} we show the validity of the fitting $\tanh$ function for $\mu(k,a)$ as measured from $k$-evolution. For both speeds of sound at some redshifts and all scales, the accuracy of the fit is of the sub-percent level. In Fig.~\ref{fitting_mu_validity2} the relative difference between fit and data from \class is shown. {For both speeds of sound at all redshifts and all scales, $\tanh$ is a good fit}. In Fig.~\ref{fig:mu_params_evolution} the evolution of fit parameters for \kev and \class data is shown.

\section{Scalar-vector-tensor decomposition and notation} \label{svt_decom}
 In this appendix we briefly discuss the scalar-vector-tensor decomposition and we introduce our notation for the metric perturbations after the decomposition.
Using the SVT decomposition \cite{Lifshitz:1945du} we can decompose $B_{i}$ into the curl-free (longitudinal) and  divergence-free (transverse) components,
\be 
B_i= B_i^{\perp} + B_i^{\parallel}  \text{     where      } \vec{\nabla} \cdot B^{\perp}=\vec{\nabla} \times  B^{\parallel}=0
\ee
 Also we can decompose the tensor perturbations analogously,
  \be 
 h_{ij} = h_{ij}^{\parallel} + h_{ij}^{\perp}+  h_{ij}^{(S)}  ~,
 \ee
 Here 
 \be
  h_{ij}^{\parallel} = \big( \nabla_i \nabla_j - \frac{1}{2} {\delta_{ij}} \nabla^2 \big) \Phi_h ~. \label{eq:beta}
 \ee
 where $\Phi_h$ \cite{Bertschinger:2001is} is a scalar and we have assumed that $h_{ij}$ is traceless, and 
  \be
  h_{ij}^{\perp} = \nabla_i h_j^{\perp}+\nabla_j h_i^{\perp} ~.
 \ee
 where $h_i^{\perp}$ is a divergenceless vector. The two degrees of freedom left in the tensor modes $h_{ij}^{(S)}$ correspond to the two polarisations of gravitational waves.\\
Fixing the gauge to Poisson gauge will remove two vector and two scalar degrees of freedom as we have the following constraints,
\be
\delta^{i j} B_{i, j}=\delta^{i j} h_{i j}=\delta^{j k} h_{i j, k}=0 ~. \label{eq:gauge}
\ee
\bibliographystyle{JHEP}

\bibliography{Ref}

\end{document}